\documentclass[9pt,twocolumn,twoside]{pnas-new}

\usepackage{bm}
\usepackage{amsmath}
\usepackage{caption}
\usepackage{subcaption}
\usepackage{standalone}

\templatetype{pnasresearcharticle}

\title{Moir\'e magnets}

\author[b,1]{Kasra Hejazi}
\author[a,1]{Zhu-Xi Luo} 
\author[a,c,1,2]{Leon Balents}

\affil[a]{Kavli Institute for Theoretical Physics, University of
  California, Santa Barbara, CA 93106-4030}
\affil[b]{Physics Department, University of
  California, Santa Barbara, CA 93106-4030}
\affil[c]{Canadian Institute for Advanced Research, Toronto, Ontario, Canada}

\leadauthor{Hejazi} 

\significancestatement{Moir\'e patterns, which result when two or more two dimensional materials with incommensurate or rotated lattices are layered together, create controllable {\em electronic} bands that have recently been shown to induce a tremendous wealth of physical phenomena. Here, we initiate the theoretical study of moir\'e patterns' influence on {\em magnetic} states of localized spins.    We construct a general formalism using continuum field theory, and present thorough analyses in twisted bilayers of antiferromagnets and also ferromagnets, which are within the experimental reach of Van der Waals materials.  Two-dimensional magnets, well-known for their strong intrinsic spin fluctuations, may serve as a new platform for moir\'e effects and open the door to a large class of novel phenomena that were once unimaginable.}

\authorcontributions{L.B. designed research, and carried out initial calculations; all authors performed research and wrote the manuscript.}

\authordeclaration{No conflicts of interest.}
\equalauthors{\textsuperscript{1} K.H., Z.L., and L.B. contributed equally to this work.}
\correspondingauthor{\textsuperscript{2}To whom correspondence should
  be addressed. E-mail: balents@kitp.ucsb.edu}

\keywords{Moir\'e $|$ Van der Waals magnets $|$ Hydrodynamics $|$ Twisted bilayer} 

\begin{abstract}
We introduce a general framework to study moir\'e structures of two-dimensional Van der Waals magnets using continuum field theory. The formalism eliminates quasiperiodicity and allows a full understanding of magnetic structures and their excitations.   In particular, we analyze in detail twisted bilayers of N\'eel antiferromagnets on the honeycomb lattice. A rich phase diagram with non-collinear twisted phases is obtained, and spin waves are further calculated. Direct extensions to zig-zag antiferromagnets and ferromagnets are also presented. We anticipate the results and formalism presented to lead to a broad range of applications to both fundamental research and experiments. 
\end{abstract}

\dates{This manuscript was compiled on \today}
\doi{\url{www.pnas.org/cgi/doi/10.1073/pnas.XXXXXXXXXX}}

\begin{document}

\maketitle
\ifthenelse{\boolean{shortarticle}}{\ifthenelse{\boolean{singlecolumn}}{\abscontentformatted}{\abscontent}}{}

\dropcap{T}he wealth of new phenomena revealed in incommensurate layered structures of graphene and other two dimensional semi-conductors and semi-metals have sparked major efforts in the study of electronic physics atop moir\'e patterns. The materials from which these structures are made, Van der Waals (VdW) solids, come in many varieties, inspiring a nascent field going well beyond graphene\cite{NovoselovReview}.  In particular, a growing family of VdW {\em magnets} are being explored both for their magnetism directly as well as for the interplay of that magnetism with electronics \cite{BurchReview}.  Two dimensional magnets are of particular interest for the fluctuation effects inherent to them.  For example, the Mermin-Wagner theorem \cite{MerminWagner} proves that a strictly two-dimensional magnet with Heisenberg or XY symmetry cannot show long-range order at any non-zero temperature. Exotic quantum phases of magnets, e.g. quantum spin liquids, are widely expected to be more prevalent in two dimensions\cite{0034-4885-80-1-016502}.

In this paper, we introduce a framework to study moir\'e structures of two dimensional magnets, under assumptions which are widely applicable and achievable in VdW systems. We present a general methodology to derive continuum models for incommensurate/twisted/strained multilayers including the effects of interlayer coupling, obviating the need to consider thousands or tens of thousands of lattice sites/spins with complicated local environments.  We illustrate the method with detailed calculations for the case of a twisted bilayer of two-sublattice N\'eel antiferromagnets on the honeycomb lattice, a situation realized
in MnPS$_3$ \cite{Ouvrard, LeFlem, Kurosawa},  MnPSe$_3$ \cite{LeFlem}, and also discuss applications to honeycomb
lattice antiferromagnets with zig-zag magnetic order (as in FePS$_3$ \cite{Kurosawa}, CoPS$_3$ \cite{Ouvrard}, NiPS$_3$ \cite{OuvrardThesis}, see \cite{Brec} for a review) and to the honeycomb lattice {\em
ferromagnet} CrI$_3$ \cite{Huang}.  We show that twisting these magnets leads to controllable emergent non-collinear spin textures (despite the fact that the parent materials all exhibit collinear ordering), and a rich spectrum of magnonic subbands. 

Now we turn to the exposition of the problem and approach, which we
illustrate as we go for the simplest case of a two-sublattice N\'eel
order on the honeycomb lattice. First we
detail the assumptions under which a continuum description is
possible.  We consider structures built from two dimensional magnets
with long range magnetic order at zero temperature, and assume that
the inter-layer exchange interactions $\sim J'$ are weak compared to
the intra-layer exchange $J$, i.e. $J'\ll J$.  Additionally, we assume
that the lattice in each layer may be regarded as a deformed version
of a parent structure shared by all layers.  Each layer $l$ is
described by a displacement field $\bm{u}_l(\bm{x})$ in Eulerian
coordinates:
\begin{equation}
  \label{eq:48}
  \bm{u}_l(\bm{x}_l) = \bm{x}_l-\bm{x}_l^{(0)},
\end{equation}
where $\bm{x}_l$ and $\bm{x}_l^{(0)}$ are the deformed and original positions,
respectively, of points in layer $l$. 
The displacement field of each layer need not
be uniform or small but its gradients should be small,
i.e. $|\partial_\mu u_{l,\nu}|\ll 1$.  For uniform
layers, this allows any long-period moir\'e structure, i.e. for which
the period of the moir\'e pattern is large compared to the magnetic
unit cell.  For two identical but twisted layers, it corresponds to
the case of a small twist angle, $\theta \ll \pi$.  This construction
is directly analogous to the procedure to build the continuum model of
twisted bilayer graphene\cite{bistritzer2011moire} following the
recent derivation in Ref.\cite{10.21468/SciPostPhys.7.4.048} which is
valid under nearly identical assumptions.

In this situation the interlayer couplings and the displacement
gradients are small perturbations on the intrinsic magnetism of the
layers, and therefore have signficant effects only at low energies.
This allows a continuum representation of the magnetism of each layer
in terms of its low energy modes: space-time fluctuations of the order
parameters.   The order parameter of the two sublattice antiferromagnet is a
N\'eel vector $\bm{N}_l$ with fixed length $|\bm{N}_l|=1$, and its low
energy dynamics for an isolated undeformed layer is described by the
non-linear sigma model with the Lagrange density
\begin{equation}
  \label{eq:49}
  \mathcal{L}_0[\bm{N}_l]= \frac{\rho }{2v^2} \left(\partial_t\bm{N}_l\right)^2 -
  \frac{\rho}{2} \left( \nabla\bm{N}_l\right)^2 + d\left(N_l^z\right)^2,
\end{equation}
where $\rho\sim J$ is the spin stiffness, $v$ is the spin-wave
velocity, and $d$ is a uniaxial anisotropy with $d>0$ signifying
Ising-like and $d<0$ XY-like anti-ferromagnetism.  For MnPS$_3$, there
is weak Ising-like anisotropy \cite{Wildes} so $0<d \ll J/A_{\rm u.c.}$
($A_{\rm u.c.}$ is the area of the 2d unit cell). Such smallness (but
not the sign) of the anisotropy is common for third row transition
metal magnets.

Next we consider the first order effects of displacement gradients
upon the intra-layer terms in \eqref{eq:49}.  As in
Ref.~\cite{10.21468/SciPostPhys.7.4.048}, such terms arise from pure
geometry -- i.e. carrying out the coordinate transformation from
$\bm{x}_l^{(0)}$ to $\bm{x}_l$ defined in \eqref{eq:48} -- and
from strain-induced changes in energetics.  Taking them together, the
leading corrections to \eqref{eq:49} are
\begin{equation}  \label{eq:50}
  \begin{aligned}
    \mathcal{L}_1[\bm{N}_l,\bm{u}_l] =& \rho(\varepsilon_{l,xx}+\varepsilon_{l,yy})
    \left[
  \frac{\delta_1}{v^2} \left(\partial_t\bm{N}_l\right)^2 -
  \delta_2  \left( \nabla\bm{N}_l\right)^2\right] \\
  & + \delta_3
\varepsilon_{l,\mu\nu} \partial_\mu \bm{N}_l \cdot \partial_\nu \bm{N}_l,
\end{aligned}
\end{equation}
where $\delta_{1,2,3}$ are dimensionless $O(1)$ constants and
$\varepsilon_{l,\mu\nu} = (\partial_\mu u_{l,\nu}+ \partial_\nu
u_{l,\mu})/2$ is the strain field in layer $l$.  For simplicity we
assumed that spin-orbit effects (e.g. anisotropy $d$) are small and
hence that deformation terms in \eqref{eq:50} are SU(2) invariant:
 anisotropic deformation terms must be small in both spin-orbit
 coupling and in displacement gradients, and hence are neglected.

 Next we turn to the inter-layer coupling terms.  By locality and
 translational symmetry, it is generally of the form
 \begin{equation}
   \label{eq:1}
   \mathcal{L}_2[\bm{N}_1,\bm{N}_2,\bm{u}_1-\bm{u}_2] = J'[\bm{u}_1-\bm{u}_2] \bm{N}_1\cdot\bm{N}_2,
 \end{equation}
 where $J'[\bm{u}]$ is a function with the periodicity of the
 undeformed Bravais lattice.   Due to the smallness of $J'$, we
 neglect corrections proportional to displacement gradients in
 \eqref{eq:1}.  Generally $J'[\bm{u}]$ can be expanded in a Fourier
 series, and well-approximated by a small number of harmonics.  We
 obtain a specific form by considering local coupling of the spin
 densities in the two layers.  Using the symmetries of the honeycomb
 lattice, the minimal Fourier expansion of the spin density
 $\mathcal{S}_l$ of a
 single layer contains
 the three minimal reciprocal lattice vectors $\bm{b}_a$,
 \begin{equation}
   \label{eq:2}
   \mathcal{S}_l(\bm{x}) = n_0\bm{N}_l \sum_{a=1}^3 \sin (\bm{b}_a \cdot\bm{x}^{(0)})= n_0\bm{N}_l \sum_{a=1}^3\sin [\bm{b}_a \cdot(\bm{x}-\bm{u}_l)],
 \end{equation}
 where $n_0$ measures the size of the ordered moment, and we define the origin $\bm{x}=\bm{0}$ at the center of a
 hexagon.  Taking the product $\mathcal{S}_1\cdot\mathcal{S}_2$ and
 applying trigonometric identities to extract the terms which vary
 slowly on the lattice scale (rapidly varying components do not
 contribute at low energy) gives the form of \eqref{eq:1}, with
 \begin{equation}
   \label{eq:3}
   J'[\bm{u}] = J' \sum_{a=1}^3 \cos (\bm{b}_a \cdot \bm{u}),
 \end{equation}
 where the constant $J'$ is proportional to the inter-layer
 exchange and $n_0^2$.  Physically, \eqref{eq:3} captures the fact
 that e.g. for intrinsically ferromagnetic exchange $J'>0$, the
preferred relative orientation of the A sublattice spins of the two
layers is parallel for AA stacking but anti-parallel for AB and BA
stackings.

The full Lagrange density $\mathcal{L} = \sum_{l=1,2}
(\mathcal{L}_0[\bm{N}_l] + \mathcal{L}_1[\bm{N}_l,\bm{u}_l]) +
\mathcal{L}_2[\bm{N}_1,\bm{N}_2,\bm{u}_1-\bm{u}_2] $ captures the low
energy physics of a bilayer with arbitrary deformations of the two
layers.  We now specialize to the case of a rigid twist of the two
layers by a relative angle $\theta$: $\bm{u}_1 = -\bm{u}_2 =
\frac{\theta}{2} \bm{\hat{z}}\times\bm{x}$.  In this case the strain
vanishes, and one finds the full Lagrangian is
\begin{equation}
  \label{eq:4}
  \mathcal{L} = \sum_l \frac{\rho }{2v^2}
  \left(\partial_t\bm{N}_l\right)^2- \mathcal{H}_{\rm cl},
\end{equation}
where
\begin{equation}
  \label{eq:5}
  \mathcal{H}_{\rm cl} = \sum_l \left[ \frac{\rho}{2} \left(
      \nabla\bm{N}_l\right)^2 - d
    \left(N_l^z\right)^2\right] - J' \Phi(\bm{x}) \bm{N}_1\cdot \bm{N}_2,
\end{equation}
is the classical energy density.  Here the coupling function
\begin{equation}
  \label{eq:6}
  \Phi(\bm{x}) = \sum_{a=1}^3 \cos (\bm{q}_a \cdot \bm{x}), 
\end{equation}
and $\bm{q}_a = \theta\bm{\hat{z}}\times \bm{b}_a$ are the reciprocal
lattice vectors {\em of the moir\'e superlattice}.  

Equations (\ref{eq:4})-(\ref{eq:6}) form the basis for an analysis of the
magnetic structure on the moir\'e scale, as well as for the magnon
excitations above them. 
The magnetic ground state is obtained as
the variational minimum of $\mathcal{H}_{\rm cl}[\bm{N}_1,\bm{N}_2]$.
Owing to the sign change of $\Phi(\bm{x})$, the problem is frustrated:
the N\'eel vectors of the two layers wish to be parallel in some
regions and antiparallel in others, forcing them to develop gradients
within the plane -- the representation in the continuum of incompletely
satisfied in-plane bonds. We find that  the optimal classical solution is coplanar but not necessarily collinear (see the Supporting Information for a complete
  weak coupling analysis), and without loss of generality we can take
the spins to lie in the x-z plane:
$\bm{N}^{\rm cl}_l = \sin\phi_l\bm{\hat{x}}+
\cos\phi_l\bm{\hat{z}}$.  The formulae are simplified by forming symmetric and antisymmetric
combinations, $\phi_s= \phi_1+\phi_2$, $\phi_a = \phi_1-\phi_2$, and
adopting dimensionless coordinates ${\sf x} = q_m \bm{x}$, with
$q_m=|\bm{q}_a|$ the moir\'e wavevector.  Then the dimensionless
energy density ${\sf H}_{\rm cl}=\mathcal{H}_{\rm
  cl}/(\frac{1}{2}\rho q_m^2)$ becomes,
up to an additive constant
\begin{equation}
  \label{eq:7}
 {\sf H}_{\rm cl} = \frac{1}{2}
  \left( |\bm{\nabla}_{\sf x}\phi_s|^2 +
    |\bm{\nabla}_{\sf x}\phi_a|^2\right) - (\alpha \hat\Phi(\sf{x})+\beta\cos\phi_s) \cos
  \phi_a.
\end{equation}
Here we introduced the dimensionless parameters
\begin{equation}
  \label{eq:10}
  \alpha = \frac{2J'}{\rho q_m^2}, \qquad \beta = \frac{2d}{\rho q_m^2},
\end{equation}
and $\hat\Phi({\sf x}) =  \sum_{a=1}^3 \cos (\bm{\hat{q}}_a \cdot
\sf{x})$, where $\bm{\hat{q}}_a=\bm{q}_a/q_m$ are unit vectors.
We can obtain partial differential equations for the phase angles by
applying calculus of variations to \eqref{eq:7}:
\begin{align}
  \label{eq:8}
  \nabla_{\sf x}^2 \phi_s & = \beta \cos \phi_a \sin \phi_s, \\
  \nabla_{\sf x}^2\phi_a & = \left(\beta \cos\phi_s +
                       \alpha \hat\Phi({\sf x})\right) \sin\phi_a.\label{eq:9}
\end{align}

We must find the solutions of the saddle point equations which minimize the integral of ${\sf H}_{cl}$.  There is always a trivial solution with $\phi_s=\phi_a=0,\pi$, which corresponds to the Ising limit of aligned or counter-aligned spins. Nontrivial solutions with potentially lower energies will be discussed in different limits below.  We first focus on the case $\beta>0$.

For $\alpha,\beta\ll 1$, corresponding to large angles, the gradient terms in the Hamiltonian dominate and the solution is nearly constant. Perturbation theory with fixed $\delta = \beta/\alpha^2$ gives a nontrivial solution
\begin{equation}\label{eq:13}
\begin{aligned}
 \phi_s &= \pi , \\
  \phi_a &= \phi_a^{(0)}-\alpha \, \sin\phi_a^{(0)} \left( \hat\Phi({\sf x}) - \Phi_0 \right) + O(\alpha^2,\beta), 
\end{aligned}
\end{equation}
with $\cos\phi_a^{(0)} = - \frac23 \, \delta$, and where $\Phi_0$ is a constant given in Sec.~C of the Supporting Information.
In this \emph{twisted} solution, $\cos\phi_a$ tends to imitate the sign of $\hat\Phi({\sf x})$, to gain energy from the potential term. This change of $\phi_a$, however, will need to be balanced against the energy penalty due to the kinetic term.
Comparing the energy of the twisted solution with the trivial one, 
we see that it has a lower energy for $\delta < \frac32$, i.e., whenever it exists; in this limit, at $\delta = \frac32$, the system undergoes a continuous transition to the collinear phase. More details can be found in the Supporting Information.

For small angles, on the other hand, $\alpha \gg 1$; in this limit, we first consider small values of $\beta$; the potential term $\alpha \hat{\Phi}(\sf{x})$ in \eqref{eq:7} dominates and the energy is minimized by choosing $\phi_a=0$ or $\pi$ almost everywhere, so that $\cos\phi_a = {\rm sign}[\hat{\Phi}(\sf{x})]$, which means the order parameter vectors in the two layers are locally parallel or antiparallel. At small values of $\beta$, it is preferred for $\phi_s$ to take a constant value and, since the total area with negative $\hat{\Phi}(\sf{x})$ is larger than the positive area, $\phi_s=\pi$ is chosen; this solution lies in the same phase as that of the twisted solution found for $\alpha \ll 1$ above, that also showed the property $\phi_s=\pi$; we call this phase \textit{twisted-s}. The twisted-s solution obviously breaks the U(1) symmetry of spin rotations about the $z$ axis of spin space, but it retains an Ising symmetry under interchanging layers and reflecting spin $N^z \rightarrow -N^z$.  One may check that $\varphi=\phi_s-\pi$ is odd under this symmetry.

Interestingly however, one can further check that in the same limit of $\alpha \gg 1$, above some order-one value of $\beta$, another twisted solution becomes more energetically favored. It belongs to what we call a \emph{twisted-a} phase, where $\phi_s$ is no longer constant and actually shows a twisted pattern similar to that of $\phi_a$, such that $\cos\phi_s$ exhibits spatial variations following those of $\hat{\Phi}(\sf{x})$ (see the Supporting Information for details).  This implies a non-zero value for $\varphi$ so that the twisted-a phase spontaneously breaks the aforementioned Ising symmetry.  The value of $\varphi \neq 0$ increases smoothly from zero on entering the twisted-a phase from the twisted-s one, consistent with the expected continuous behavior of an Ising transition (treated at mean-field level by the saddle point analysis). 

One can finally study the $\alpha,\beta \gg 1$ limit. The twisted-a solution in this limit, which is the lowest energy nontrivial solution, requires $\phi_s$ along with $\phi_a$ to also have the values $0$ or $\pi$ almost everywhere so that $\cos\phi_s$ matches the sign of $\hat{\Phi}(\sf{x})$; this means simply that the vectors $\bm{N}_l$ align or counter-align along the $\pm\bm{\hat{z}}$ axis almost everywhere. The order parameter rotations occur in a narrow domain wall centered on the zeros of
$\hat{\Phi}(\sf{x})$, i.e.~forming a closed almost circular loop in the
middle of the unit cell.    This domain wall costs an energy
proportional to its length;  as a result of this energy penalty, the twisted-a phase gives way to the collinear phase when the energy gain from the twist is exceeded by the
domain wall energy. In order to study this competition, we assume that such transition occurs when $\beta\gg\alpha$, which we later check is self-consistent.  The widths
${\sf w}_{a}$ and ${\sf w}_s$ (in dimensionless distance normal to the domain wall) 
over which $\phi_a$ and $\phi_s$ wind are determined by the balance of the gradient terms and the corresponding potential terms. This gives ${\sf w}_{a/s} \sim 1/\sqrt{\beta}$ in this limit and an energy cost per unit length of the wall of ${\sf E} \sim \sqrt{\beta}$.
 Now the bulk
energy gain of the twisted state is simply proportional to $\alpha$,
so we obtain the result that the twist collapses when
$\sqrt{\beta} \gtrsim \alpha$.  This treatment is valid since under these conditions
$\beta/\alpha \gtrsim \alpha \gg 1$ (we did not determine the multiplicative order one
constant in this inequality).  Note that the transition between the collinear and twisted-a phase is a ``level crossing" between two distinct and disconnected saddle points; consequently it corresponds to a {\em first order} transition, and the first derivatives of the energy are discontinuous across this boundary.  A tricritical point separates the two continuous transitions from this first order one.

\begin{figure}[t]
    \centering
    \includegraphics[scale=0.4]{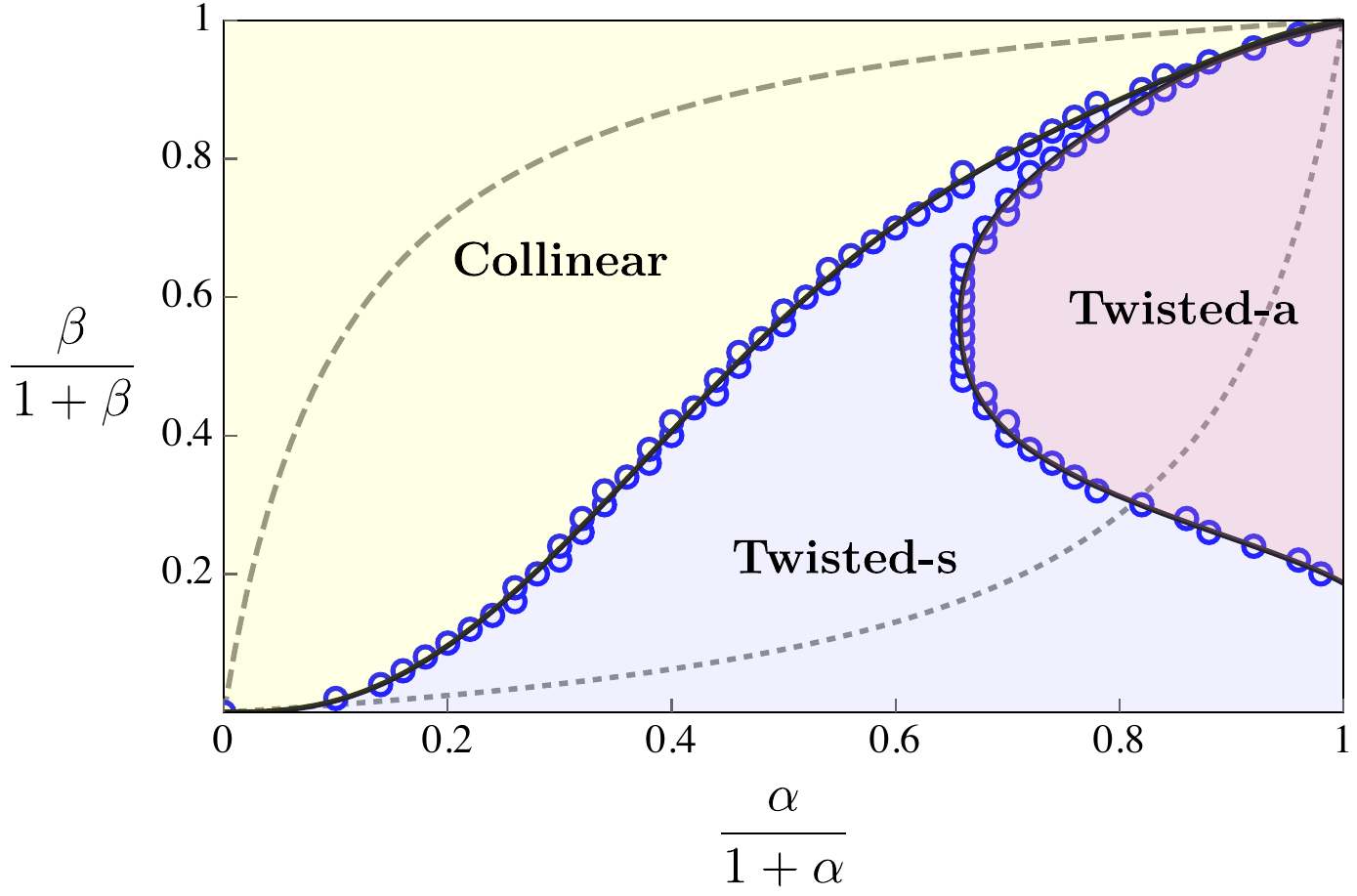}
    \caption{The phase diagram with respect to the normalized dimensionless parameters $\alpha/(1+\alpha)$ and $\beta/(1+\beta)$, respectively. In the collinear phase, the Neel vectors of the two layers are constant and either aligned or counter-aligned. In the twisted-s phase, $\phi_s=\pi$ while the sign of $\cos\phi_a$ modulates with that of $\hat{\Phi}(\sf x)$. In the twisted-a phase, both signs of $\cos\phi_s$ and $\cos\phi_a$ follow that of $\hat{\Phi}(\sf x)$. The twisted-s phase terminates at $(0.88,0.94)$ near the right top corner of the diagram. The dashed line shows $\beta=10\alpha$ while the dotted line corresponds to $\beta=0.1\alpha$. For $\beta=\alpha$, the corresponding line would be the diagonal one connecting the left bottom and right top corners.}
    \label{fig:phases}
\end{figure}

\begin{figure*}[ht]
\centering
\hspace{-1cm}
\centering
\begin{subfigure}[c]{0.32\textwidth}
    \centering
    \includegraphics[width=\textwidth]{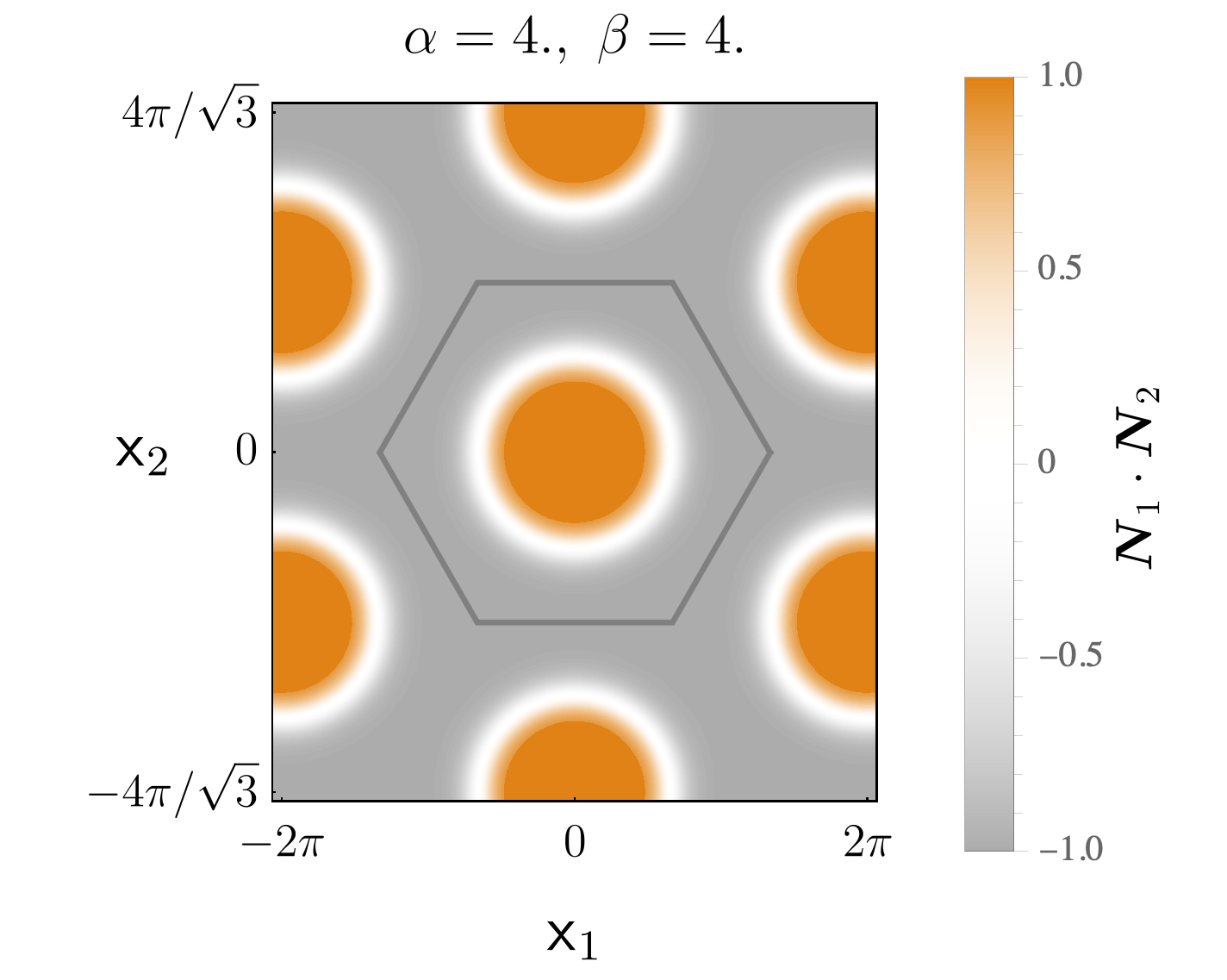}
    \caption{\label{fig:twisted_solutions_afm_a}}
\end{subfigure}\hspace{0.8cm}
\begin{subfigure}[c]{0.16\textwidth}
    \centering
    \includegraphics[width=\textwidth]{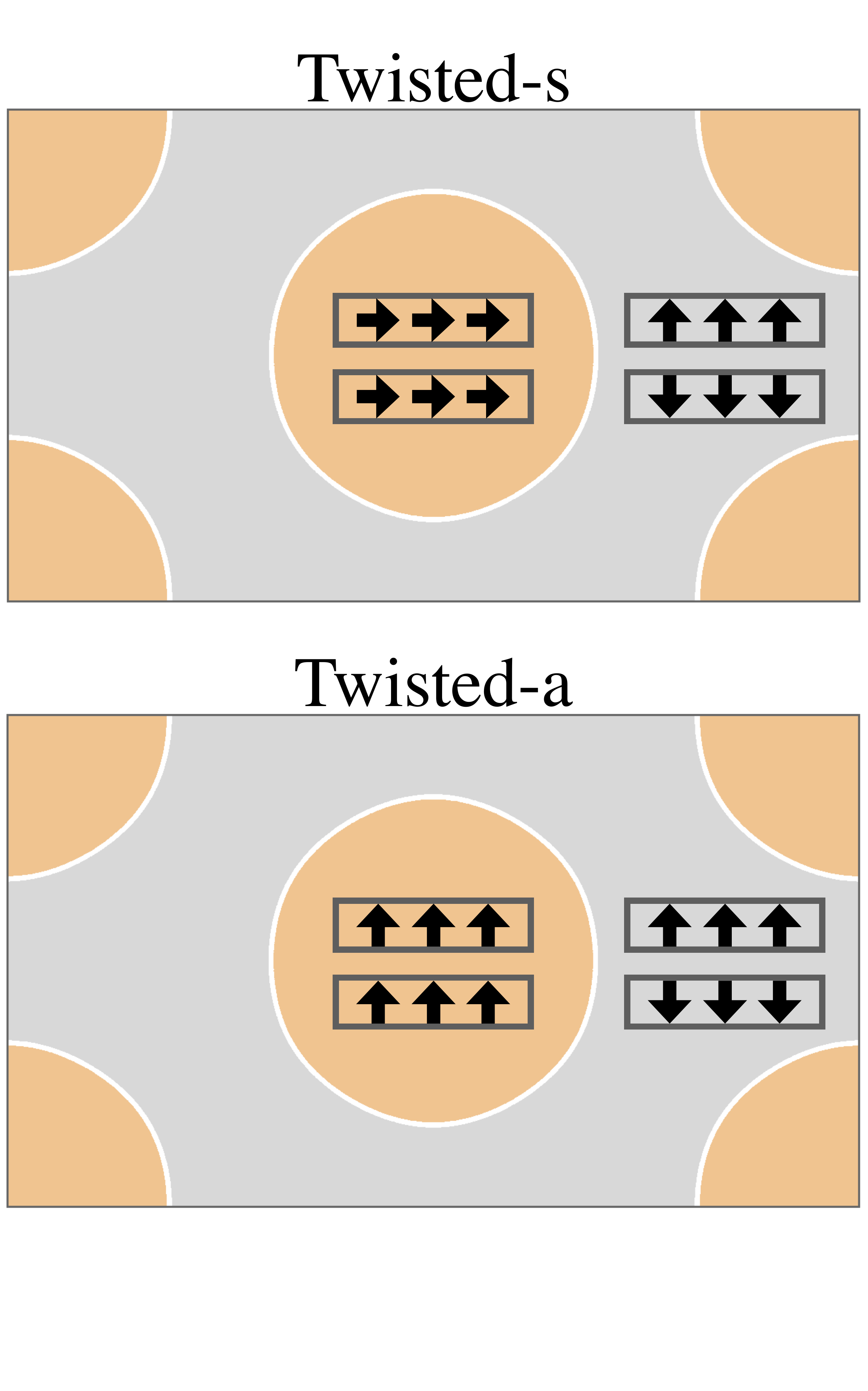}
    \caption{\label{fig:twisted_solutions_afm_b}}
\end{subfigure} \hspace{1.2cm}
\begin{subfigure}[c]{0.44\textwidth}
    \centering
        \centering
    \includegraphics[scale=0.3]{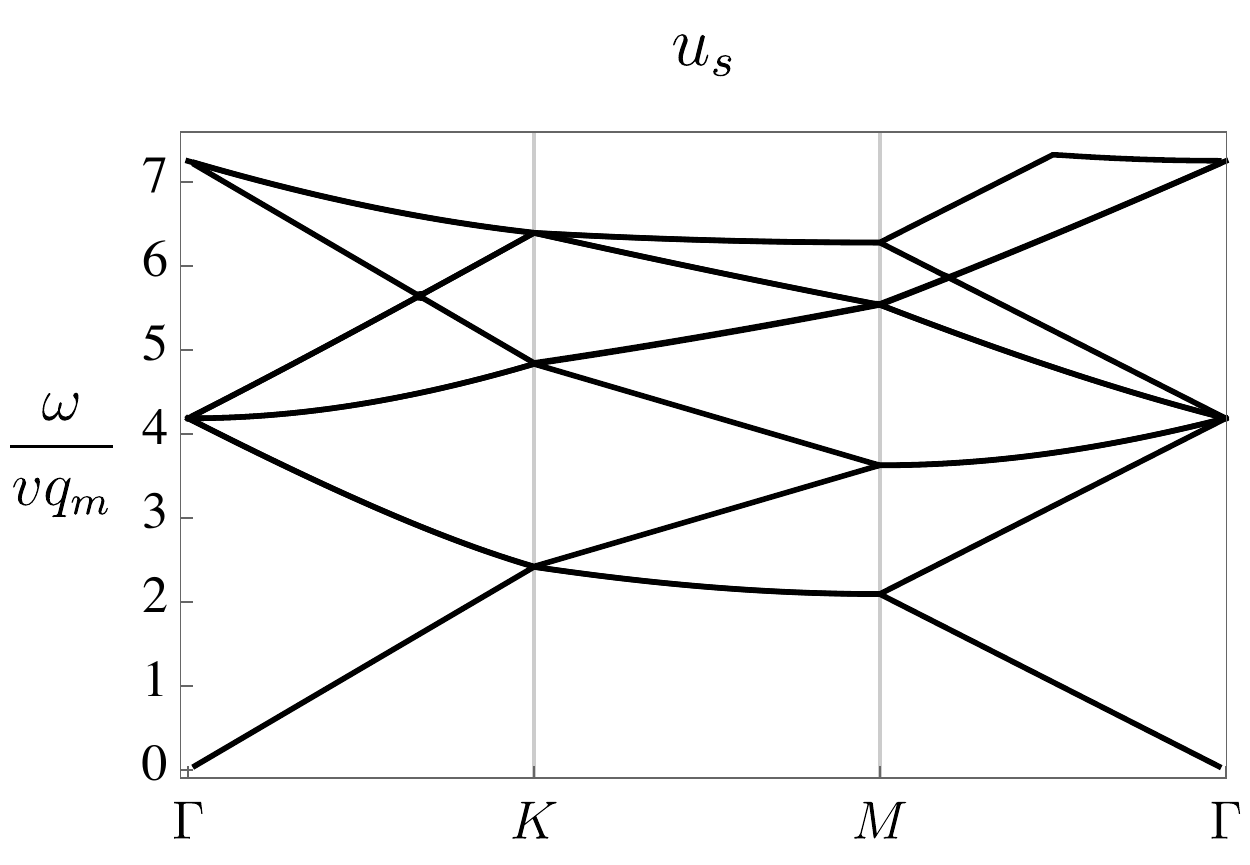}
    \includegraphics[scale=0.3]{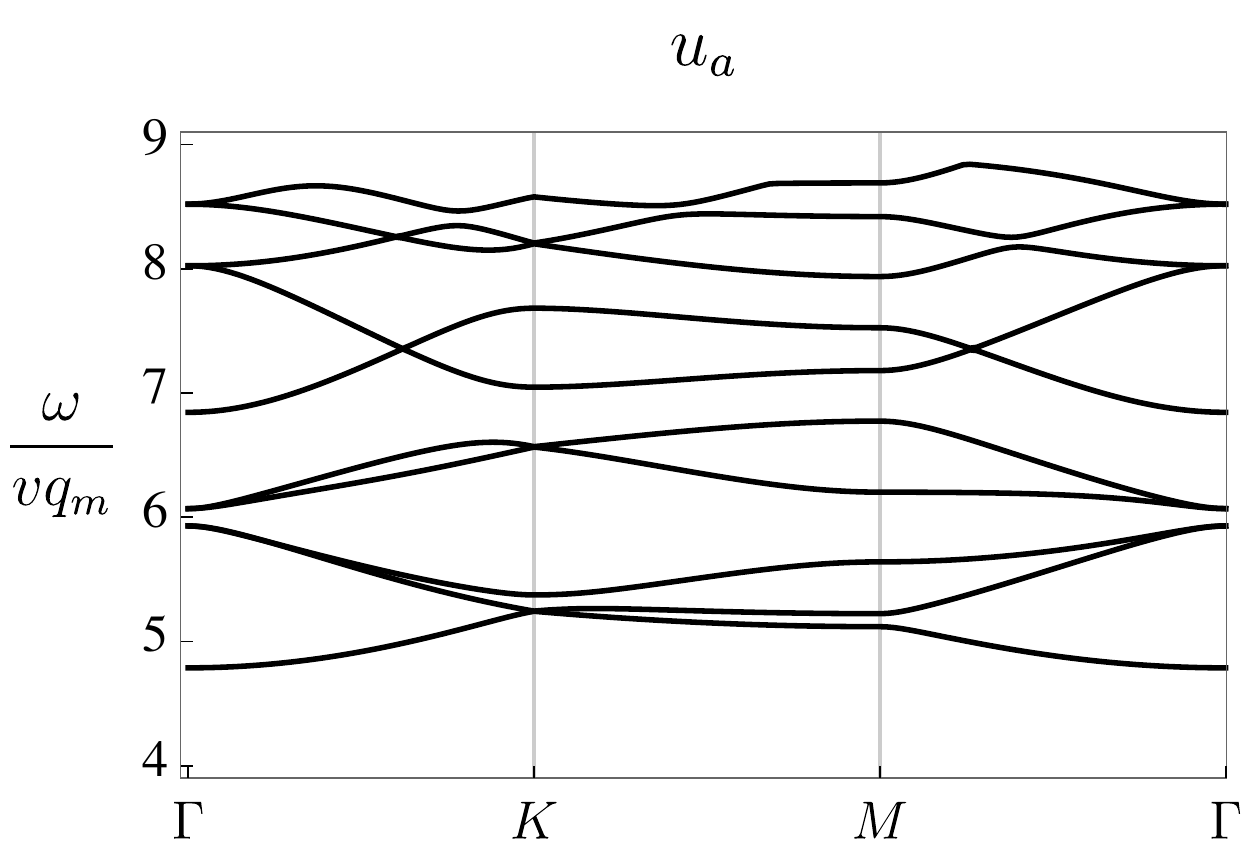}
    \includegraphics[scale=0.3]{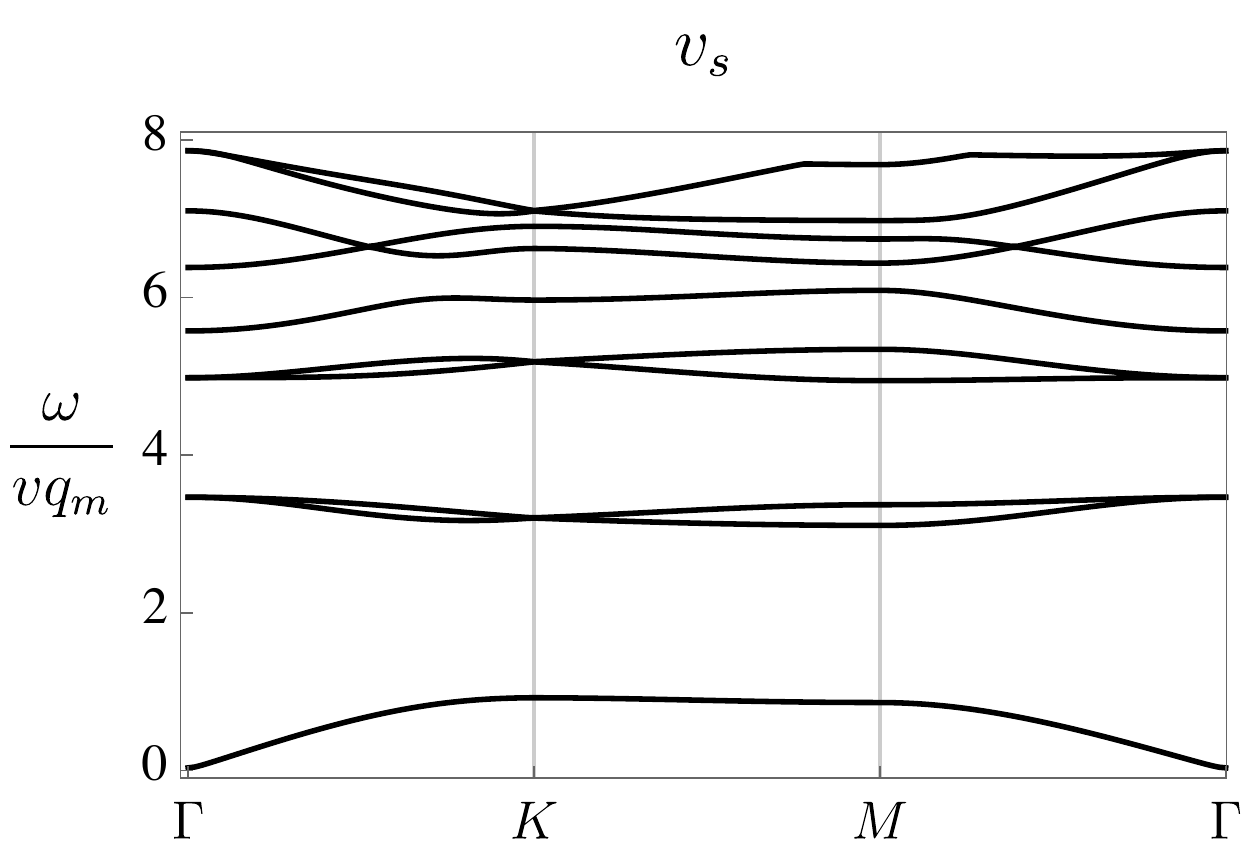}
    \includegraphics[scale=0.3]{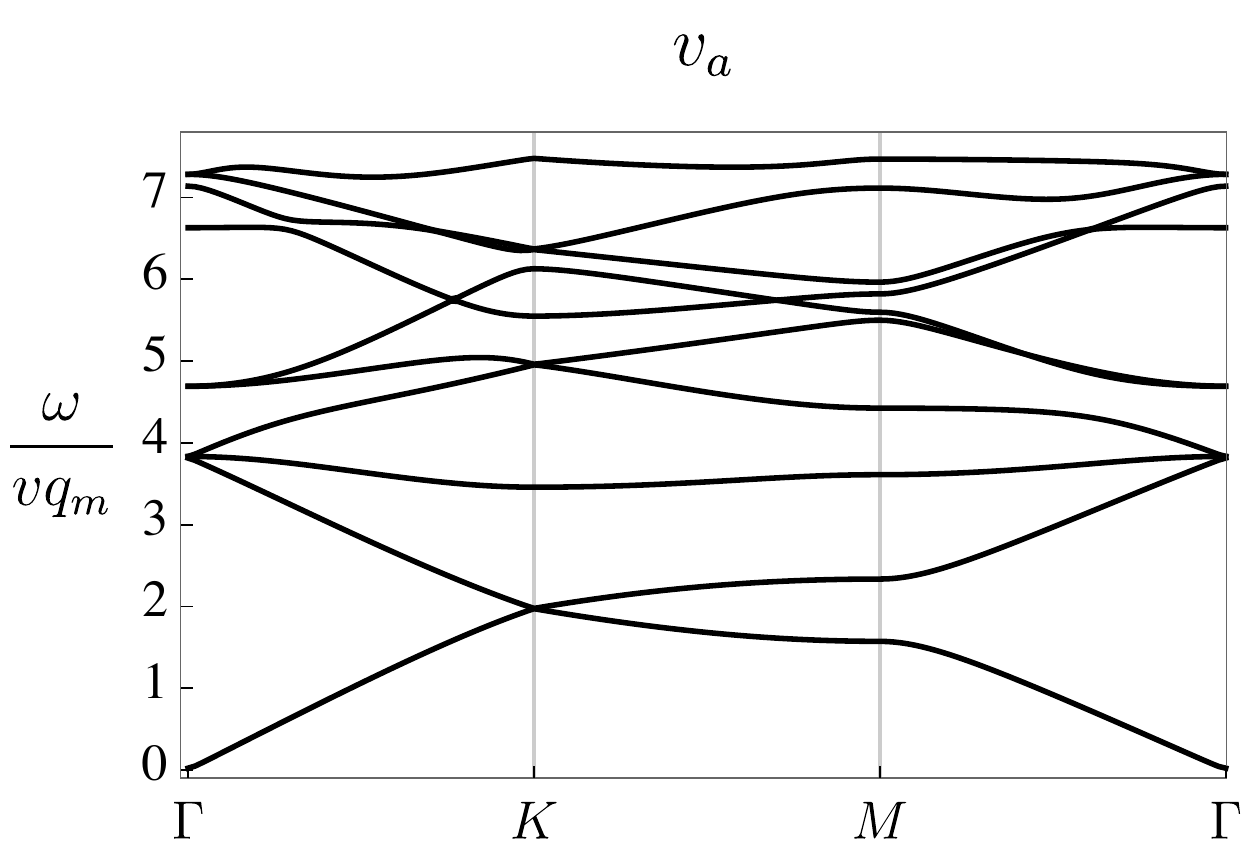}
    \caption{\label{fig:magnon}}
\end{subfigure}
\caption{(a) Spatial configurations of $\bm{N}_1 \cdot \bm{N}_2 = \cos \phi_a$. It can be seen that $\cos \phi_a$ is either $+1$ or $-1$ almost everywhere, based on the sign of $\hat\Phi({\sf x})$. The choice $\alpha = 4 , \beta = 4$ corresponds to the twisted-a phase; the same quantity, i.e.~$\cos \phi_a$ looks very similar if $\beta$ is lowered even into the twisted-s phase while keeping $\alpha$ fixed.  
(b) A schematic diagram of the spatial dependence of the orientation of the order parameters (not actual spins) in the two layers, a vertical (horizontal) arrow denotes out-of-plane (in-plane) order parameter orientation; the brown and gray areas show regions with positive and negative values for $\hat\Phi({\sf x})$. The main difference is that in the brown region, the twisted-s phase shows in plane orientation and the twisted-a phase shows out-of-plane orientation. 
(c) The lowest ten magnon bands at $\alpha=2$ for the four branches, in the isotropic case ($d=0$) of two coupled two-sublattice antiferromagnets.}
\end{figure*}

To summarize, we find three different phases for $\beta>0$, two of which correspond to twisted solutions. The transition between the two twisted phases happens at some $\beta$ of order $1$, when the phase boundary is crossed in the large $\alpha$ limit; the twisted phases collapse on the other hand when $\beta >
\frac{3}{2}\alpha^2$ in the $\alpha,\beta \ll 1$ limit (twisted-s to Ising transition) and when $\beta \gtrsim \alpha^{2}$ in the $\alpha,\beta \gg 1$ limit (twisted-a to Ising transition). 
We sketch a phase diagram in Fig. \ref{fig:phases} from the numerical solution to \eqref{eq:8}, \eqref{eq:9} in Fig. \ref{fig:phases}, which is consistent with and in fact interpolates between the perturbative and strong-coupling analyses above. The dashed and dotted lines in the figure show examples of paths with a fixed ratio $\beta/\alpha=d/J'$; this ratio is determined by the material, but one can tune the twisting angle $\theta$ to move along the lines, and consequently enter/leave different phases.
Remarkably, for a fixed $d/J'$, {\em the twisted-a state is always stabilized for sufficiently small $\theta$}; this can be understood by noting that $\beta/\alpha$ is invariant when $\theta$ changes as mentioned above, and thus decreasing the twist angle, increases $\alpha$ linearly with $\beta$ forming a straight line in the $\alpha-\beta$ plane (different from the axes in Fig.~\ref{fig:phases}), but the twisted-a phase, when $\alpha,\beta \gg 1$, is separated from the collinear phase by a $\beta \sim \alpha^2$ relation and from the twisted-s phase by $\beta\sim \mathrm{const.}$; thus the above mentioned straight line lies between these two phase boundaries at sufficiently small $\theta$. Plots of the real space configurations of the ground states in the two twisted solutions are presented in Figures \ref{fig:twisted_solutions_afm_a} and \ref{fig:twisted_solutions_afm_b}.

Once the minimum energy saddle point is obtained, the full Lagrangian allows for calculation of the magnon spectrum.  We define 
\begin{equation}
  \label{eq:u_v_definition}
   \bm{N}_l = \sqrt{1-u_l^2 -v_l^2} \bm{N}_l^{\rm cl}({\sf x}) + u_l
   \bm{u}_l({\sf x}) +
  v_l \bm{v}_l({\sf x}),
\end{equation}
where $\bm{u}_l = \cos \phi_l({\sf x})\bm{\hat{x}} - \sin  \phi_l({\sf
  x})\bm{\hat{z}}$ and $\bm{v}_l = \bm{\hat{y}}$ complete a {\em
  spatially dependent}
orthonormal basis such that $\bm{\hat{u}}\times\bm{\hat{v}} =
\bm{N}_l^{\rm cl}$ at every $\bm{x}$.  The fluctuations about the
classical solution are described by space-time dependent fields
$u_l({\sf x},t),v_l({\sf x},t)$.  Inserting \eqref{eq:u_v_definition} into the \eqref{eq:4}, expanding to quadratic order in the fluctuations and finding the Euler-Lagrange equations for $u_{s/a} = u_1\pm
u_2,v_{s/a}=v_1\pm v_2$, one obtains linear
wave equations for four branches of excitations.  For simplicity, we
present the results for $d=\beta=0$ (see Supporting Information for the general
  result), in which case the four modes decouple immediately
\begin{equation}
  \label{eq:16}
  -\partial_t^2 u_{s/a} = v^2 q_m^2 \hat{\sf D}_{u,s/a} u_{s,a},
  \qquad  -\partial_t^2 v_{s/a} = v^2 q_m^2 \hat{\sf D}_{v,s/a} v_{s,a},
\end{equation}
with the linear operators
\begin{align}
  \label{eq:17}
  \hat{\sf D}_{u,s} = & - \nabla_{\sf x}^2, \\
  \hat{\sf D}_{u,a} = & - \nabla_{\sf x}^2+\alpha
                              \hat{\Phi}({\sf x}) \cos\phi_a,
                        \nonumber \\
  \hat{\sf D}_{v,s} = & - \nabla_{\sf x}^2- \frac{1}{4}
                       |\bm{\nabla}_{\sf x}\phi_a|^2 -
                       \frac{\alpha}{2}\hat\Phi({\sf x})
                              \left(1-\cos\phi_a\right), \nonumber \\
    \hat{\sf D}_{v,a} = & - \nabla_{\sf x}^2 - \frac{1}{4}
                       |\bm{\nabla}_{\sf x}\phi_a|^2 +
                       \frac{\alpha}{2}\hat\Phi({\sf x}) \left(1+\cos\phi_a\right)\nonumber.
\end{align}
Taking $u_{s/a}({\sf x},t) = e^{i\omega t} u_{s/a}({\sf x})$, we
obtain eigenvalue problems such that the magnon frequencies are
($vq_m$ multiplied by) the
square roots of the eigenvalues of the $\hat{\sf D}$ operators.  These
eigenvalue problems have the form of continuum non-relativistic
Schr\"odinger-Bloch problems and therefore can be solved using the
Bloch ansatz to find an infinite series of magnon bands. When $\alpha$ is large, as the potential terms in the above equations become alternated deep wells and hard walls, which confine the magnons to either of the two domains. This leads to the flattening of magnon bands in branches $u_a$ and $v_s$.
Fig. \ref{fig:magnon} shows the lowest magnon bands when $\alpha$ is at intermediate value. There are three gapless Goldstone modes in the $u_s$,  $v_s$ and $v_a$ branches, which correspond to the three generators of the $O(3)$ group. 

Finally, we comment on the case of $d<0$ in brief, where the anisotropy term favors the spins to lie in the XY-plane. The corresponding equations of motion resemble those of the isotropic case, i.e., $\phi_s$ tends to be uniform everywhere, while $\cos\phi_a$ imitates the sign of $\hat{\Phi}({\sf x})$ due to the interlayer exchange, leading to twisted configurations. More details can be found in the Supporting Information.

\subsubsection*{Zig-zag antiferromagnet}
\label{sec:zig-zag-antif}

Having described the case of the N\'eel antiferromagnet in detail, we
give further results more succinctly for other types of 2d magnets.
The materials FePS$_3$, CoPS$_3$, and NiPS$_3$ all have the same
lattice structure as MnPS$_3$ but exhibit ``zig-zag'' magnetic order.   It
is a collinear magnetic order which doubles the unit cell.  There are
three possible ordering wavevectors: the M points at the centers of
the edges of the moir\'e Brillouin zone, which are half reciprocal lattice vectors,
$\bm{b}_a/2$, with $a=1,2,3$.  The spin density (analogous
to \eqref{eq:2}) therefore contains three order parameter ``flavors'',
$\bm{N}_a$:
\begin{equation}
  \label{eq:spin_zigzag}
  \mathcal{S}_l(\bm{x}) = n_0 \sum_{a=1,2,3} \bm{N}_{l,a} \sin \left[\frac{1}{2}\bm{b}_a \cdot\left(\bm{x}-\bm{u}_l\right)\right].
\end{equation}
Here in a zig-zag state, just a single one of the three $\bm{N}_a$
vectors is non-zero: this describes three possible spatial
orientations of the zig-zag chains of aligned spins.  Proceeding as before, we obtain the
effective classical Hamiltonian in the form
\begin{align}
  \label{eq:15}
  & \mathcal{H}^{\rm zig-zag}_{\rm cl} =  \sum_{a,l} \left[\frac{\rho}{2}
  (\bm{\nabla}\bm{N}_{a,l})^2 +\frac{\tilde{\rho}}{2}
  (\bm{\hat{q}}_a\cdot\bm{\nabla}\bm{N}_{a,l})^2\right] \\ &+ \sum_l V[\bm{N}_{1,l},\bm{N}_{2,l},\bm{N}_{3,l}]
   - \frac{J'}{2} \sum_a \bm{N}_{a,1}\cdot \bm{N}_{a,2} \cos
  \left(\frac{\bm{q}_a\cdot\bm{x}}{2}\right).\nonumber
\end{align}
Here $\rho,\tilde{\rho}$ are two stiffness constants, and
$V$ is a potential which may be taken in the form
\begin{equation}
  \label{eq:18}
  V[\bm{N}_1,\bm{N}_2,\bm{N}_3] = u (\sum_a |N_a|^2-1)^2 + v
  \sum_{a>b} |N_a|^2 |N_b|^2 - d\sum_a (N_a^z)^2,
\end{equation}
with $u,v>0$ to model the energetic preference for a single non-zero
stripe orientation, and $d$ as before to tune anisotropy.  

\begin{figure}[t]
    \centering
    \includegraphics[scale=0.4]{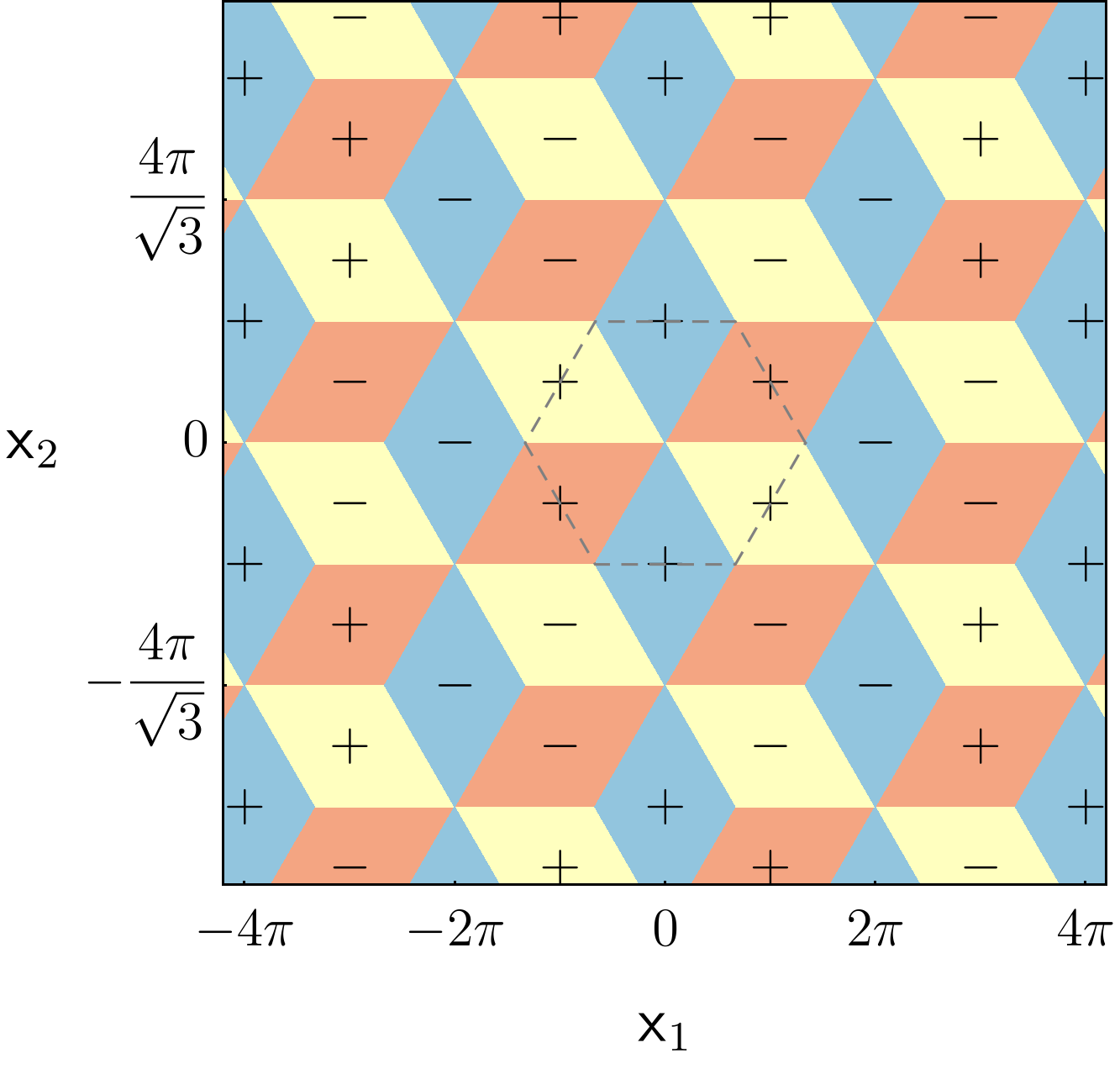}
    \caption{The real space tiling with six possible domains which appear in a twisted bilayer of zig-zag antiferromagnets in the strong coupling (small angle) limit.  The colors show which flavor of the order parameter is nonzero in each domain, while the $\pm$ signs show the relative sign between the order parameter in the two layers. The dashed hexagon shows a moir\'e structural unit cell.}
    \label{fig:zigzag_pattern}
\end{figure}

\eqref{eq:15} gives a continuum model to determine the magnetic
ordering texture for arbitrary twist angles.  The most important
difference from the two-sublattice antiferromagnet is that here each
spatial harmonic couples to a single ``flavor'', while in the former
case, \eqref{eq:spin_zigzag}, the single flavor of order parameter couples to
the sum of harmonics.  While we do not present a general solution, we
note immediate consequences in the strong coupling limit,
$J'\gg \rho q_m^2,\tilde{\rho}q_m^2$.  In this situation, for each
$\bm{x}$ we must choose the {\em largest} harmonic, i.e. the $a$ which
maximizes $|\cos \left(\frac{\bm{q}_a\cdot\bm{x}}{2}\right)|$, and
then take
$\bm{N}_{a,1}={\rm sign}[J' \cos
\left(\frac{\bm{q}_a\cdot\bm{x}}{2}\right)]\bm{N}_{a,2}$ and
$N_{a',l}=0$ for $a'\neq a$.  Remarkably, the result is a tiling of
six possible zig-zag domains which evokes a ``dice lattice'', as shown
in Fig.~\ref{fig:zigzag_pattern}.  Narrow domain walls separate these regions.

\subsubsection*{Twisted ferromagnet}
\label{sec:twisted-ferromagnet}

Na\"ively, twisting a homo-bilayer of ferromagnets is relatively
innocuous.  However, experiments and theory \cite{mcguire2015coupling,song2019switching,wang2018very,huang2018electrical,sivadas2018stacking,soriano2019interplay,jiang2019stacking} for CrI$_3$ have indicated
that the interlayer exchange has a strength and sign that depends upon
the displacement between neighboring layers.  This can be directly
incorporated into a continuum model following our methodology.  

\begin{figure}[!t]
\centering
\centering
\begin{subfigure}[b]{0.154\textwidth}
    \centering
    \includegraphics[width=\textwidth]{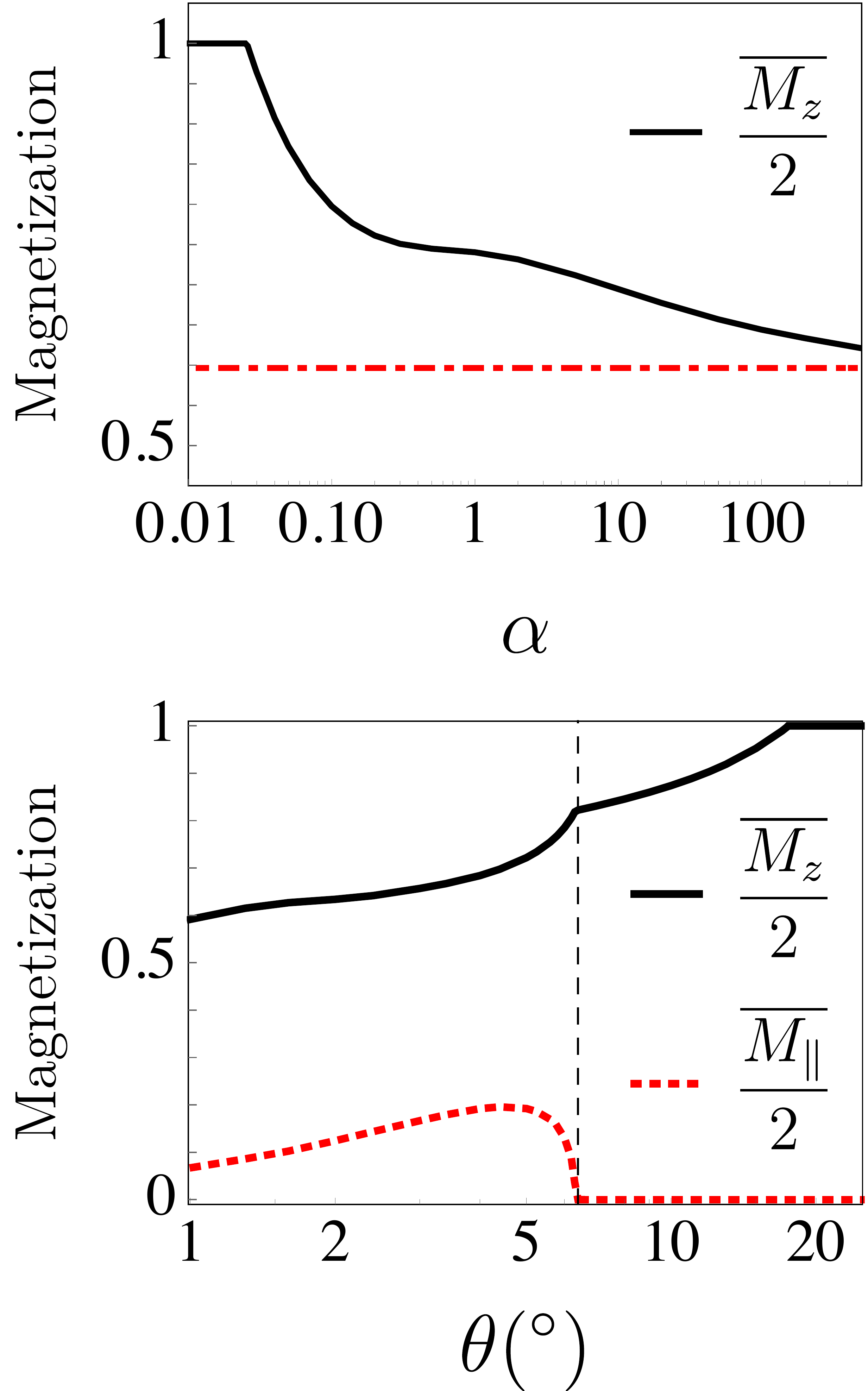}
    \caption{\label{fig:twisted_solutions_fm_a}}
\end{subfigure} \hspace{0.2cm}
\begin{subfigure}[b]{0.31\textwidth}
    \centering
    \includegraphics[width=\textwidth]{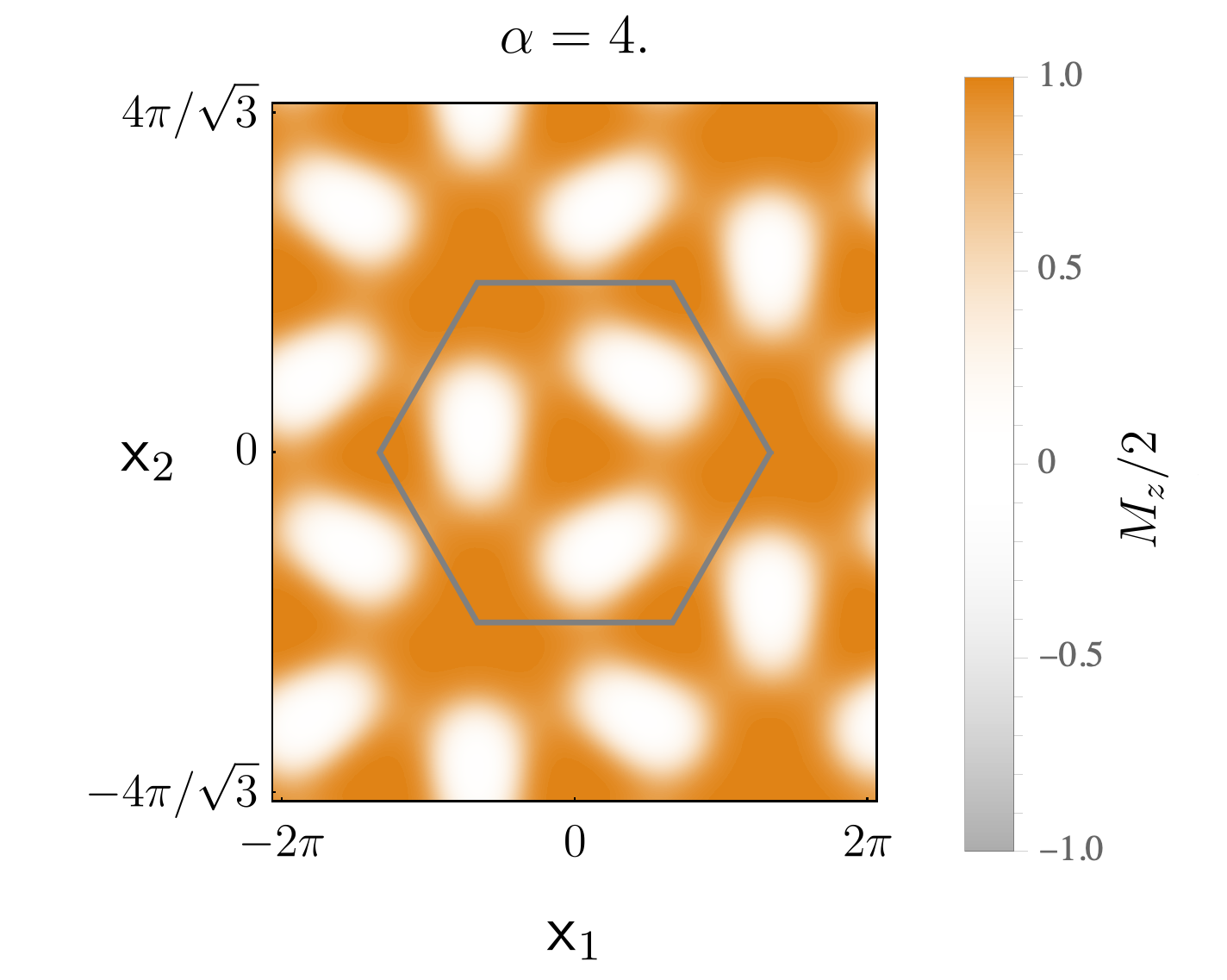}
    \caption{\label{fig:twisted_solutions_fm_b}}
\end{subfigure}
\caption{ (a) Top: the average value of the $z$ component of the sum of the two layers' spins for CrI$_3$, when the anisotropy paramter is taken to be positive and infinitesimal. A continuous transition from the collinear phase to the twisted phase occurs at $\alpha = 0.025 $. This phase is analogous to the twisted-s phase discussed previously. The total area in which $\hat\Phi({\sf x}) > 0$ is shown with a dashed red line here as the limiting value of $\frac{\overline{M_z}}{2}$ for very large $\alpha$. Bottom: the average value of the $z$ and in-plane components of the total magnetization calculated with physical parameters chosen as discussed in the main text for CrI$_3$; in particular, the anisotropy is nonzero here. At $\theta = 17.5^{\circ}$, a transition from collinear to twisted-s phase occurs at which point $\overline{M_z}$ starts to be nonzero. Moreover, a transition to the twisted-a phase occurs at $\theta = 6.4^{\circ}$, with $\overline{M_{\parallel}}$ starting to be nonzero for smaller angles.
(b) Spatial profile of local magnetization $\frac{M_z}{2} = \frac12 \left(M_{1,z} + M_{2,z}\right)$ for a twisted solution in CrI$_3$. There are large regions in real space with a net magnetization, while other regions have vanishingly small net magnetization.}
\label{fig:twisted_solutions_fm}
\end{figure}

To this end, for a general twisted bilayer of a ferromagnetic material with the above property, one can use the energy functional shown in \eqref{eq:5} with minimal modifications: \textit{i}) the N\'eel vectors $\bm{N}_l$ should be everywhere replaced by the uniform magnetization $\bm{M}_l$, since in fact each layer is ferromagnetic within itself and \textit{ii}) the function $\Phi(\bm{u}_1 - \bm{u}_2)$ takes a more complicated form.  The latter may be determined from the dependence of the interlayer exchange of untwisted layers on a uniform interlayer displacement.  For the case of CrI$_3$, we have extracted the stacking dependent interlayer exchange data from the first principle calculations in Ref. \cite{sivadas2018stacking}.
Similar to the case of twisted antiferromagnets, a variational analysis of \eqref{eq:5} can be performed, which leads to the same set of Euler-Lagrange equations, i.e.~\eqref{eq:8} and \eqref{eq:9}. In order to simplify the analysis, we will only consider an infinitesimal $\beta$ here; its effect is to fix the value of $\phi_s = 0$ for CrI$_3$ as discussed in the Supporting Information. The effects of non-zero $\beta$ can also be studied in a way similar to the previous case.  The mathematical problem is then to obtain the functional form of $\phi_a(\bm{x})$ and its dependence upon $\alpha$.   In the ferromagnetic case, the Fourier expansion of $\hat{\Phi}(\sf{x})$ generally has a nonzero constant term, which dominates the solution at small $\alpha$. In the case of CrI$_3$, the constant term is small and ferromagnetic, thus $\phi_a = 0$ is chosen for small $\alpha$.   However, if other harmonics of $\hat{\Phi}({\sf x})$ are strong enough, a \emph{twisted} solution starts to appear at a finite value of $\alpha$ with a lower energy. As in the antiferromagnetic case, $\cos\phi_a$ shows spatial modulations imitating the changes of $\hat{\Phi}({\sf x})$ in this twisted solution.   This property of the twisted solution is most visible in the large $\alpha$ limit, where the kinetic energy penalty is least important: one observes then domains with $\cos\phi_a = \mathrm{sgn} \left[ \hat{\Phi}({\sf x}) \right]$, separated by narrow domain walls. For a detailed analysis of the above statements in the case of CrI$_3$, see the Supporting Information. A plot of the average magnetization in the system is shown in Fig.~\ref{fig:twisted_solutions_fm_a} with a transition form collinear to twisted phase at finite $\alpha$. Unlike the antiferromagnets discussed above, there is a finite interval of twist angles where the collinear phase exists even with infinitesimal anisotropy parameter $\beta$. Also a plot of the spatial configuration of a twisted solution is presented in Fig.~\ref{fig:twisted_solutions_fm_b}; it shows that there are large regions in real space with maximal magnetization while at the same time there are also other regions exhibiting zero magnetization.

\subsubsection*{Conclusion}
In this work, we have considered moir\'e two dimensional magnets and in particular the twisted bilayers of Van der Waals magnetic materials. We have developed a low energy formalism in the continuum and studied in detail three different examples of twisted bilayers: antiferromagnetic, zig-zag antiferromagnetic and ferromagnetic. Remarkably, a rich phase diagram is obtained as one varies the twist angle and material parameters; there are interesting \emph{twisted} ground state solutions comprising long-wavelength non-collinear magnetic textures. Such spatial patterns can potentially be observed in experiments, where the twist angle control adds to the tunability of the system. Furthermore, at small twist angles in the non-collinear phases, certain spin waves also exhibit interesting features such as flattening dispersion curves. 

Material-wise, MnPS$_3$ has $\beta\approx 4.54\alpha$, and the system is in the collinear phase for generic twist angles. The ratio was derived using $d=\Delta^2/ \left( 12J A_{\text{u.c.}} \right)$ and $J'=J'_{\text{exp}}S^2/ \left( 2A_{\text{u.c.}} \right)$, where the intralayer exchange $J$, interlayer coupling $J'_{\text{exp}}$ and the magnon gap $\Delta$ are extracted from \cite{Wildes}. On the other hand, for CrI$_3$, we have derived $\beta \approx 0.62 \alpha$ using the intralayer exchange and anisotropy parameter as given in Ref.~\cite{chen2018topological}, and the interlayer exchange data as given in Ref.~\cite{sivadas2018stacking}. A plot of average magnetization for CrI$_3$ in the perpendicular and parallel directions for which the above parameters are used is presented in Fig.~\ref{fig:twisted_solutions_fm_a} in the bottom panel; it can be seen that at large angles, the system is in the collinear phase, but the twisted-s phase ($\phi_s = 0$) starts to be preferred at $\theta = 17.5^{\circ}$; upon further decreasing the angle, starting at $\theta = 6.4^{\circ}$, the twisted-a phase becomes the ground state. This shows that in a twisted bilayer of CrI$_3$, it is reasonable to expect both of the twisted phases to be realized in experimentally accessible settings.

The present methodology can be utilized with minimal modifications in further analyses of other moir\'e systems in the vast collection of possible bilayer magnetic materials. Their magnetic properties as well as their interplay with the electronic/transport properties could be the subject of future studies. Given the extremely fruitful research done in the field of moir\'e electronic systems, one can anticipate that the magnetic moir\'e systems could play the role of a new platform where novel exciting physics could be pursued.

\bigskip

\matmethods{
We take some space in this section to explain our numerical manipulations. 

\begin{itemize}
    \item In order to find the ground states, one needs to solve \eqref{eq:8} and \eqref{eq:9} simultaneously. We have done so by two different methods: first solving the equations in real space by the use of overdamped dynamics, i.e.~adding fictitious time derivatives of $\phi_a$ and $\phi_s$ to the equations and running the time evolution; a final configuration with zero time derivative ensures that the equations are satisfied. The second method is solving the equations in the Fourier representation iteratively by starting from a well-chosen simple guess; most of the time $\phi_a({\sf x})=\phi_s({\sf x})=\pi/2$ is a suitable initial seed. The results from the above two approaches agree completely.

\item In figure \ref{fig:phases}, the phase boundaries can be extracted by observing the changes of behavior of $\cos\phi_s$ and $\cos\phi_a$. We first solve the equations of motion \eqref{eq:8} and \eqref{eq:9} for various combinations of $\alpha$ and $\beta$, and plot the corresponding functions $\cos\phi_s$ and $\cos\phi_a$ in real space. In the collinear phase, both are constants. Fixing $\beta$ while increasing $\alpha$ from zero, $\cos\phi_a$ will begin to have spatial variation at the phase boundary between the collinear and the twisted-s phase. On the other hand, if one fixes $\alpha$ while increasing $\beta$ from zero, the $\cos\phi_s$ will start from a constant in the twisted-s phase and begin to have spatial variations once it crosses the phase boundary and enters the twisted-a phase. 

\item As for figure \ref{fig:magnon}, the spin waves are obtained from the Bloch ansatz $u_{s/a}({\sf x})=\hat{u}_{s/a}({\sf x})e^{i{\sf k}\cdot {\sf x}},$ and similarly for $v_{s/a}$. The variables in equations (\ref{eq:17}) thus become $\hat{u}_{s/a}$, $\hat{v}_{s/a}$ with the substitutions $\nabla_{\sf x}^2\rightarrow (\nabla_{\sf x}+i{\sf k})^2$ and $-\partial_t^2\rightarrow \omega^2$. Discretizing the Moir\'e unit cell, 
the linear operators become  
large matrices and can subsequently be diagonalized using Mathematica to find the magnon bands. 

\item The interlayer exchange for CrI$_3$ is extracted from Fig.~2b of Ref.~\cite{sivadas2018stacking}, where the dependence upon displacement is presented along two special lines. The interlayer exchange is a periodic function with the same period as that of the monolayer lattice; thus, a Fourier series for the interlayer exchange (a constant along with the lowest five harmonics) in the two dimensional space is assumed which induces one dimensional functions on the above two special lines; one can fix the Fourier coefficients by comparing the given functional form and the one induced by the two dimensional Fourier series. The equations are solved using the same method described above. 
\end{itemize}

All codes are available upon request.
}

\showmatmethods{} 

\acknow{We thank Andrea Young for useful discussions.  This work was supported by the Simons Collaboration on Ultra-Quantum Matter, grant 651440 from the Simons Foundation (LB,ZL), and by the DOE, Office of Science, Basic Energy Sciences under Award No. DE-FG02-08ER46524 (KH).}

\showacknow{}

\bibliography{moire}

\maketitle

\SItext

\subsection{Linear wave analysis with anisotropy}

In this section, we present the magnon spectrum when the anisotropy is present, and separately discuss the cases  of Ising anisotropy ($\beta>0$) and XY anisotropy ($\beta<0$).

\subsubsection*{Ising Anisotropy}

Following the parametrization in equation (15), 
The Hamiltonian near the saddle point $\mathbf{N}_l^{\text{cl}}$ is
\begin{equation}
\begin{split}
{\sf H}=& (\nabla_{\sf x} \phi_1)^2+(\nabla_{\sf x} \phi_2)^2-\alpha \hat{\Phi}({\sf x})\cos(\phi_1-\phi_2)-\beta (\cos^2\phi_1+\cos^2\phi_2)\\
&\left.+(\nabla_{\sf x} u_1)^2+(\nabla_{\sf x} u_2)^2 +(\nabla_{\sf x} v_1)^2+(\nabla_{\sf x} v_2)^2-v_1^2(\nabla_{\sf x} \phi_1)^2-v_2^2(\nabla_{\sf x} \phi_2)^2 \right. \\
& \left.-\frac{1}{2}\alpha \hat{\Phi}({\sf x}) \left[2v_1v_2+\cos(\phi_1-\phi_2)(2u_1u_2-u_1^2-u_2^2-v_1^2-v_2^2)\right]\right.\\
& +\beta (u_1^2 \cos 2\phi_1+u_2^2\cos2\phi_2+v_1^2\cos^2\phi_1+v_2^2\cos^2\phi_2)+\dots,
\end{split}
\end{equation}
where the first line is the 0th order contribution, and the dots in the end represent higher-order terms in $u$ and $v$. Switching to the symmetric and anti-symmetric basis, this becomes
\begin{equation}
\begin{split}
{\sf H}=& {\sf H}_{cl}+\frac{1}{2}\left[(\nabla_{\sf x} u_s)^2+(\nabla_{\sf x} u_a)^2+(\nabla_{\sf x} v_s)^2+(\nabla_{\sf x} v_a)^2\right]\\
& -\frac{1}{8}\left\{ (v_a^2+v_s^2)[(\nabla_{\sf x} \phi_a)^2+(\nabla_{\sf x} \phi_s)^2]+4v_sv_a\nabla_{\sf x} \phi_a \nabla_{\sf x} \phi_s\right\}\\
& + \frac{\alpha}{4}\hat{\Phi}({\sf x}) \left[- v_s^2(1-\cos\phi_a)+v_a^2(1+\cos\phi_a)+2u_a^2\cos\phi_a\right]\\
& +\frac{\beta}{4}\left[(v_s^2+v_a^2)+(2u_s^2+2u_a^2+v_s^2+v_a^2)\cos\phi_s\cos\phi_a-(4u_su_a+2v_sv_a)\sin\phi_s\sin\phi_a\right]
\end{split}
\end{equation}
where ${\sf H}_{cl}$ is defined in equation (10) of the main text.

For the second order terms of the Hamiltonian, we go to the Lagrangian by
\begin{equation}
{\sf L}_2=\frac{1}{2v^2}\left(|\partial_t u_s|^2+|\partial_t u_a|^2+|\partial_t v_s|^2+|\partial_t v_a|^2\right)-{\sf H}_2, 
\end{equation}
which leads to the following coupled linear wave equations for $u, v$'s:
\begin{equation}
\begin{split}
& \partial_t^2 u_s=v^2 q_m^2 \left[ \nabla_{\sf x}^2 u_s -\beta (\cos\phi_s \cos\phi_a u_s-\sin\phi_s\sin\phi_au_a) \right],\\
& \partial_t^2 u_a=v^2 q_m^2 \left[ \nabla_{\sf x}^2 u_a -\alpha \hat{\Phi}({\sf x}) \cos\phi_a u_a-\beta (\cos\phi_s\cos\phi_a u_a-\sin\phi_s\sin\phi_a u_s)\right],\\
& \partial_t^2 v_s=v^2 q_m^2[\nabla_{\sf x}^2 v_s +\frac{1}{4}\left(v_s (\nabla_{\sf x} \phi_s)^2+v_s(\nabla_{\sf x} \phi_a)^2 +2v_a\nabla_{\sf x}\phi_s\nabla_{\sf x}\phi_a\right)+\frac{\alpha}{2} \hat{\Phi}({\sf x}) (1-\cos\phi_a)v_s\\
& \quad \quad \quad \quad-\frac{\beta}{2}(v_s+v_s \cos\phi_s\cos\phi_a-v_a\sin\phi_s\sin\phi_a)],\\
& \partial_t^2 v_a=v^2 q_m^2[\nabla_{\sf x}^2 v_a +\frac{1}{4}\left(v_a (\nabla_{\sf x} \phi_s)^2+v_a(\nabla_{\sf x} \phi_a)^2 +2v_s\nabla_{\sf x}\phi_s\nabla_{\sf x}\phi_a\right)-\frac{\alpha}{2} \hat{\Phi}({\sf x}) (1+\cos\phi_a)v_s\\
& \quad \quad \quad \quad-\frac{\beta}{2}(v_a+v_a \cos\phi_s\cos\phi_a-v_s\sin\phi_s\sin\phi_a)].\\
\end{split}
\end{equation}

Taking $u_{s/a}({\sf x},t)=e^{i\omega t} u_{s/a}({\sf x})$ and using the Bloch ansatz (with ${\sf k}=\mathbf{k}/q_m$ being the dimensionless quasi-momentum vector), 
\begin{equation}
u_{s/a}({\sf x})=\hat{u}_{s/a}({\sf x})e^{i{\sf k}\cdot {\sf x}}, \quad v_{s/a}({\sf x})=\hat{v}_{s/a}({\sf x})e^{i{\sf k}\cdot {\sf x}},
\end{equation}
we obtain 
\begin{equation}
\begin{split}
& \omega^2 \hat{u}_s=-v^2 q_m^2 \left[ (\nabla_{\sf x}+i{\sf k})^2 \hat{u}_s -\beta (\cos\phi_s \cos\phi_a \hat{u}_s-\sin\phi_s\sin\phi_a\hat{u}_a) \right],\\
& \omega^2 \hat{u}_a=-v^2 q_m^2 \left[ (\nabla_{\sf x}+i{\sf k})^2 \hat{u}_a -\alpha \hat{\Phi}({\sf x}) \cos\phi_a \hat{u}_a-\beta (\cos\phi_s\cos\phi_a \hat{u}_a-\sin\phi_s\sin\phi_a \hat{u}_s)\right],\\
& \omega^2 \hat{v}_s=-v^2 q_m^2[(\nabla_{\sf x}+i{\sf k})^2 \hat{v}_s +\frac{1}{4}\left(\hat{v}_s (\nabla_{\sf x} \phi_s)^2+\hat{v}_s(\nabla_{\sf x} \phi_a)^2 +2\hat{v}_a\nabla_{\sf x}\phi_s\nabla_{\sf x}\phi_a\right)+\frac{\alpha}{2} \hat{\Phi}({\sf x}) (1-\cos\phi_a)\hat{v}_s\\
& \quad \quad \quad \quad-\frac{\beta}{2}(\hat{v}_s+\hat{v}_s \cos\phi_s\cos\phi_a-\hat{v}_a\sin\phi_s\sin\phi_a)],\\
& \omega^2 \hat{v}_a=-v^2 q_m^2[(\nabla_{\sf x}+i{\sf k})^2 \hat{v}_a +\frac{1}{4}\left(\hat{v}_a (\nabla_{\sf x} \phi_s)^2+\hat{v}_a(\nabla_{\sf x} \phi_a)^2 +2\hat{v}_s\nabla_{\sf x}\phi_s\nabla_{\sf x}\phi_a\right)-\frac{\alpha}{2} \hat{\Phi}({\sf x}) (1+\cos\phi_a)\hat{v}_a\\
& \quad \quad \quad \quad-\frac{\beta}{2}(\hat{v}_a+\hat{v}_a \cos\phi_s\cos\phi_a-\hat{v}_s\sin\phi_s\sin\phi_a)].\\
\end{split}  
\label{eq:AnisMagnon}
\end{equation}

In the collinear phase, the Neel vectors in the two layers are uniform and point either to the $+\hat{z}$ or the $-\hat{z}$ direction. Namely, there are four possible combinations that are degenerate in energy: $(\phi_s,\phi_a)=(0,0),$  $(2\pi,0),$ and $(\pi,\pi),$ $(\pi,-\pi)$. The last two of them satisfying $\cos\phi_s=\cos\phi_a=-1$ have the same discrete symmetry as that of the twisted-s phase, i.e.~a simultaneous spin reflection $N_z \to -N_z$ and layer exchange. For this type of solutions, \eqref{eq:AnisMagnon} reduces to a pair of degenerate equations for $(u_s, v_a)$ and $(u_a, v_s)$:
\begin{equation}
\begin{split}
& \omega^2 \hat{u}_s=-v^2 q_m^2 [(\nabla_{\sf x}+i{\sf k})^2-\beta]\hat{u}_s, \\
& \omega^2 \hat{u}_a=-v^2 q_m^2 [(\nabla_{\sf x}+i{\sf k})^2+\alpha \hat{\Phi}({\sf x})-\beta]\hat{u}_a.
\end{split}
\label{eq:collinear}
\end{equation}
For the other two solutions with $\cos\phi_s=\cos\phi_a=1$, \eqref{eq:AnisMagnon} reduces to a pair of degenerate equations for $(u_s, v_s)$ and $(u_a, v_a)$ instead. The $u_s$ and $u_a$ modes satisfy the same set of equations as in \eqref{eq:collinear} upon substituting $\alpha\rightarrow -\alpha$. We will choose the $\cos\phi_s=\cos\phi_a=-1$ case below for concreteness.

\begin{figure}[!h]
    \centering
    \begin{tikzpicture}
    \node at (0,0) {\tiny Collinear:};
    \node at (0,-1.1) {};
    \end{tikzpicture}
    \includegraphics[scale=0.3]{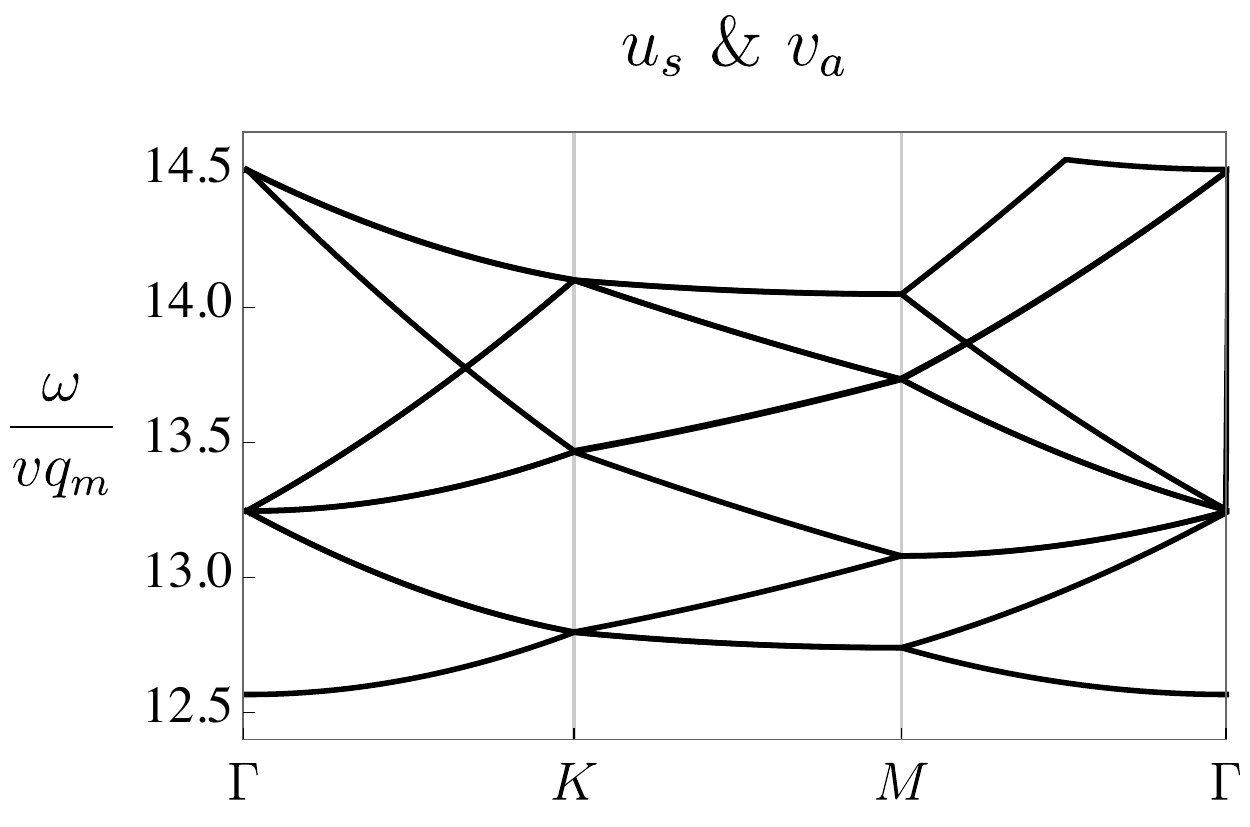}
    \includegraphics[scale=0.3]{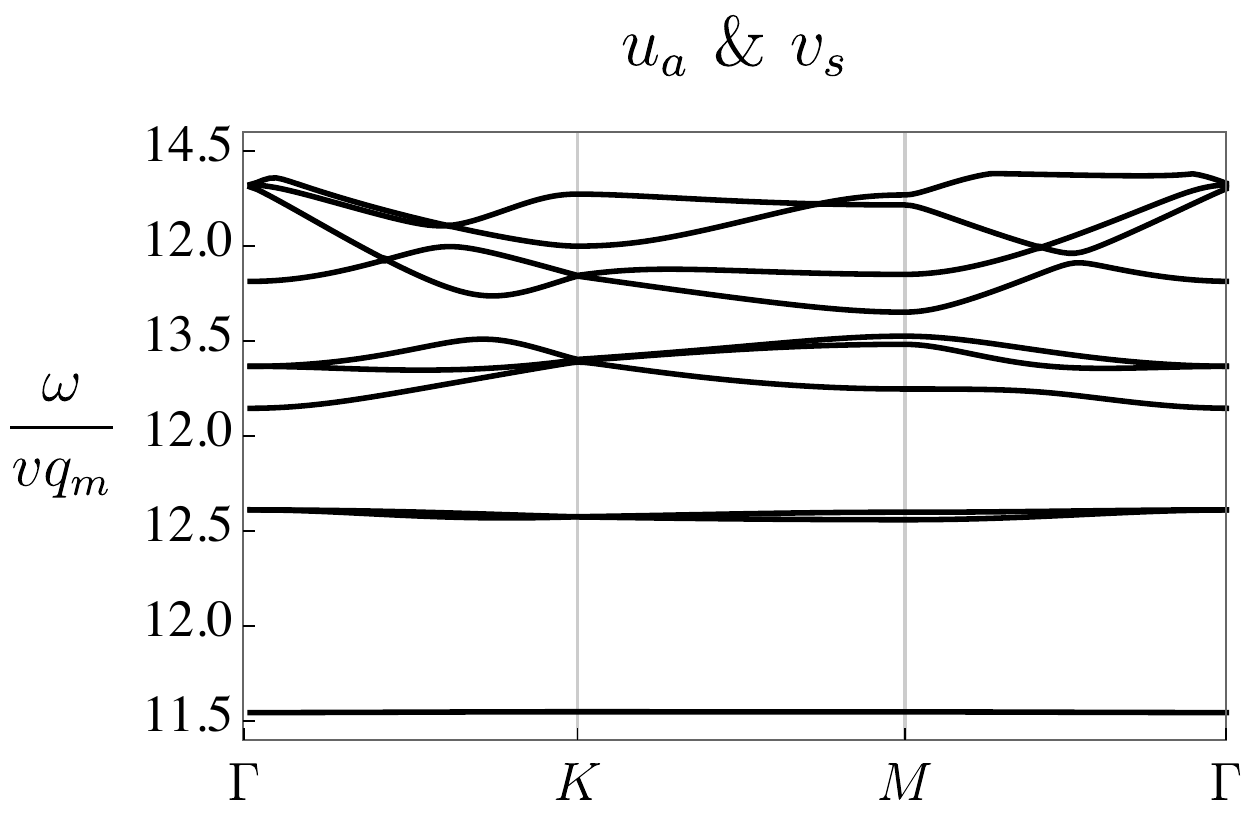}\\
    \begin{tikzpicture}
    \node at (0,0) {\tiny Twisted-a:};
    \node at (0,-1.1) {};
    \end{tikzpicture}
    \includegraphics[scale=0.3]{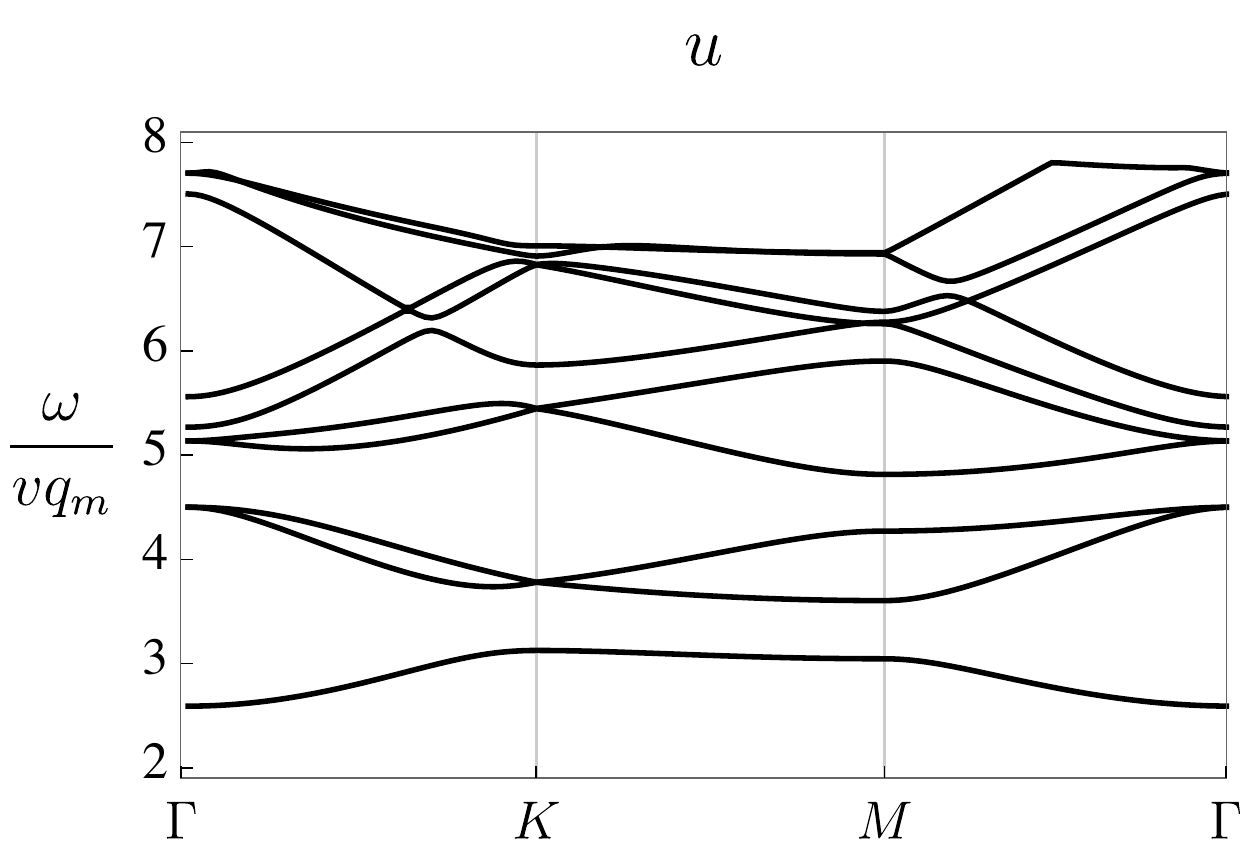}
    \includegraphics[scale=0.3]{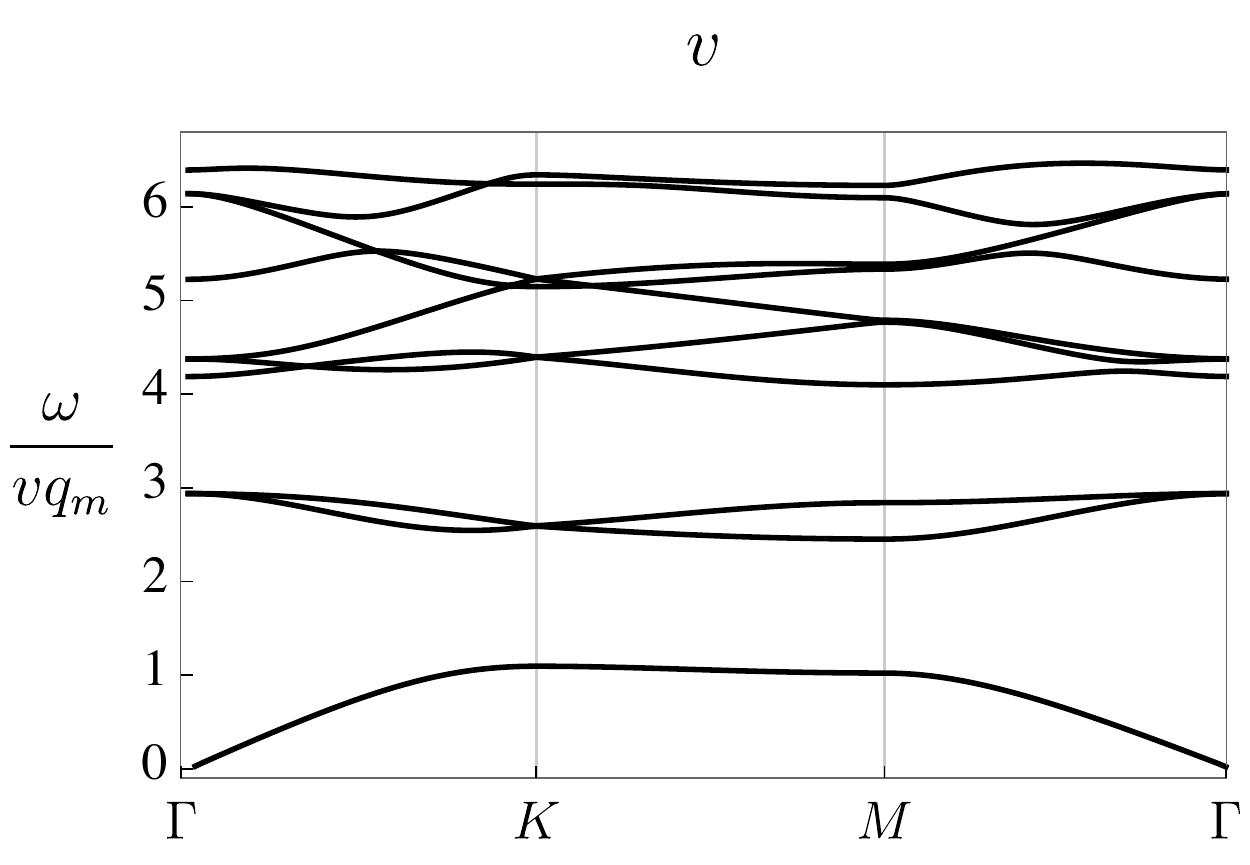}\\
    \begin{tikzpicture}
    \node at (0,0) {\tiny Twisted-s:};
    \node at (0,-1.1) {};
    \end{tikzpicture}
    \includegraphics[scale=0.3]{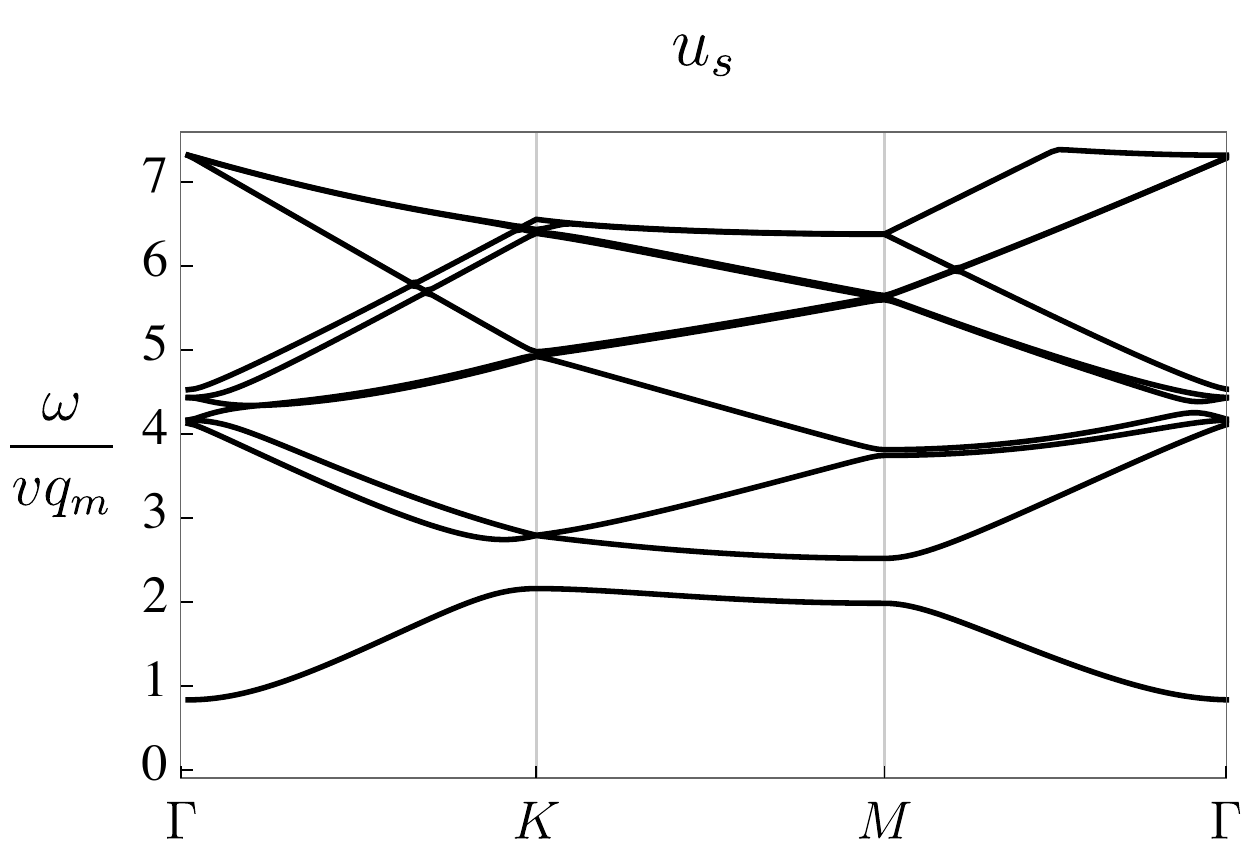}
    \includegraphics[scale=0.3]{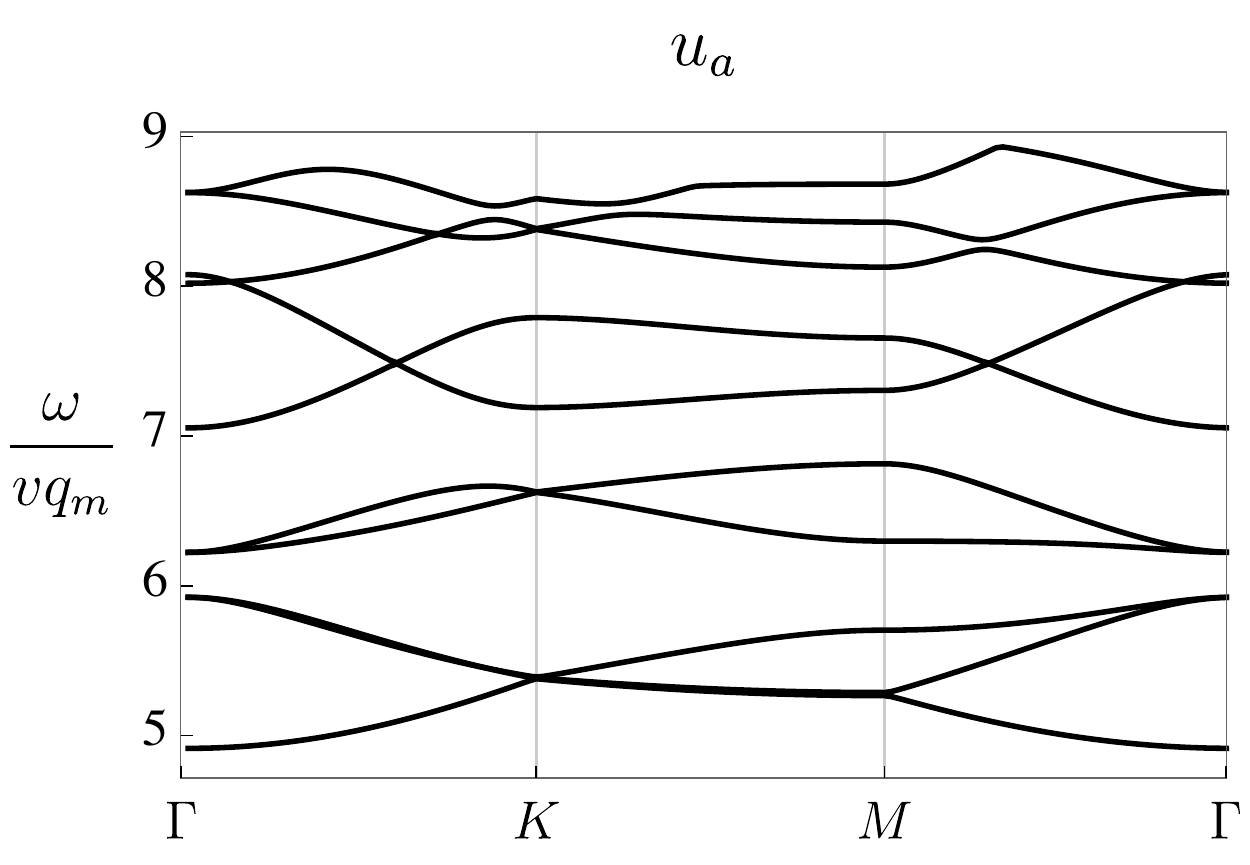}
    \includegraphics[scale=0.3]{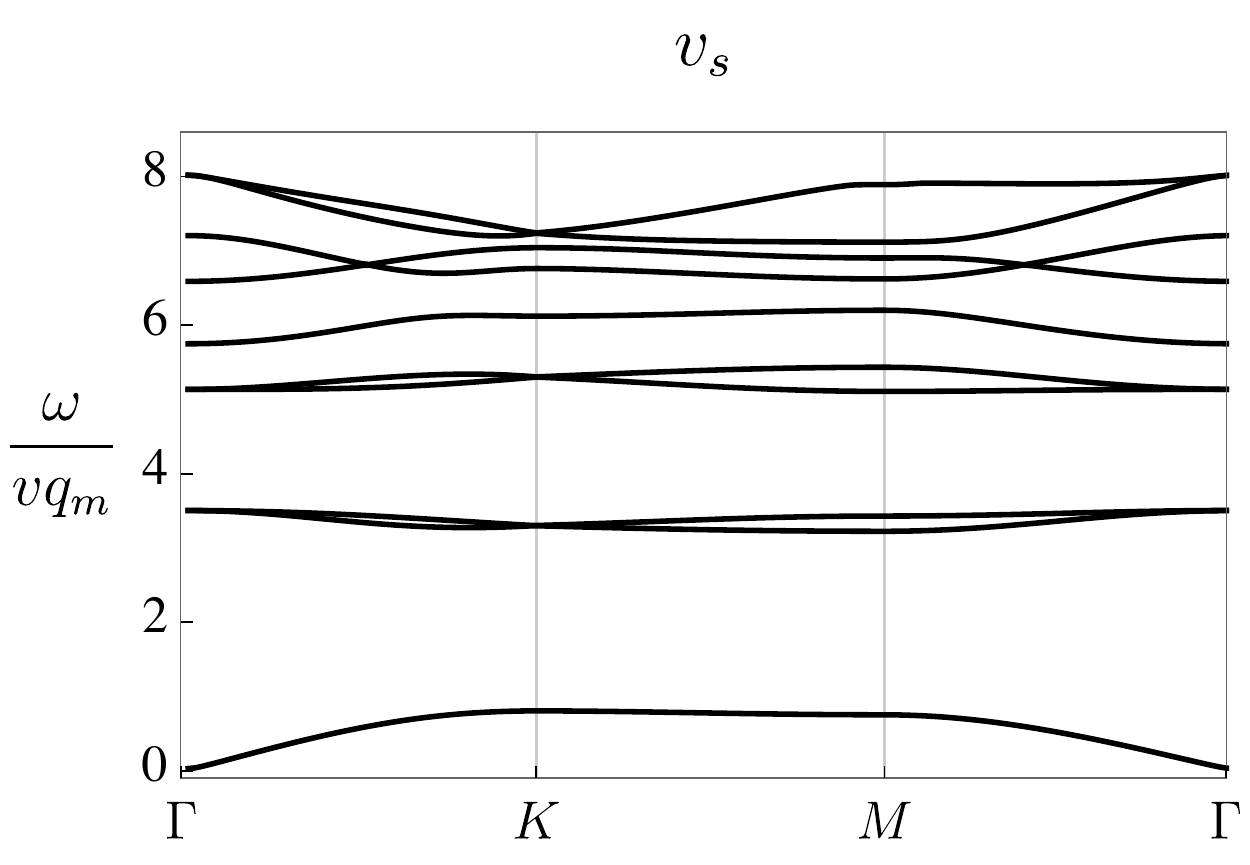}
    \includegraphics[scale=0.3]{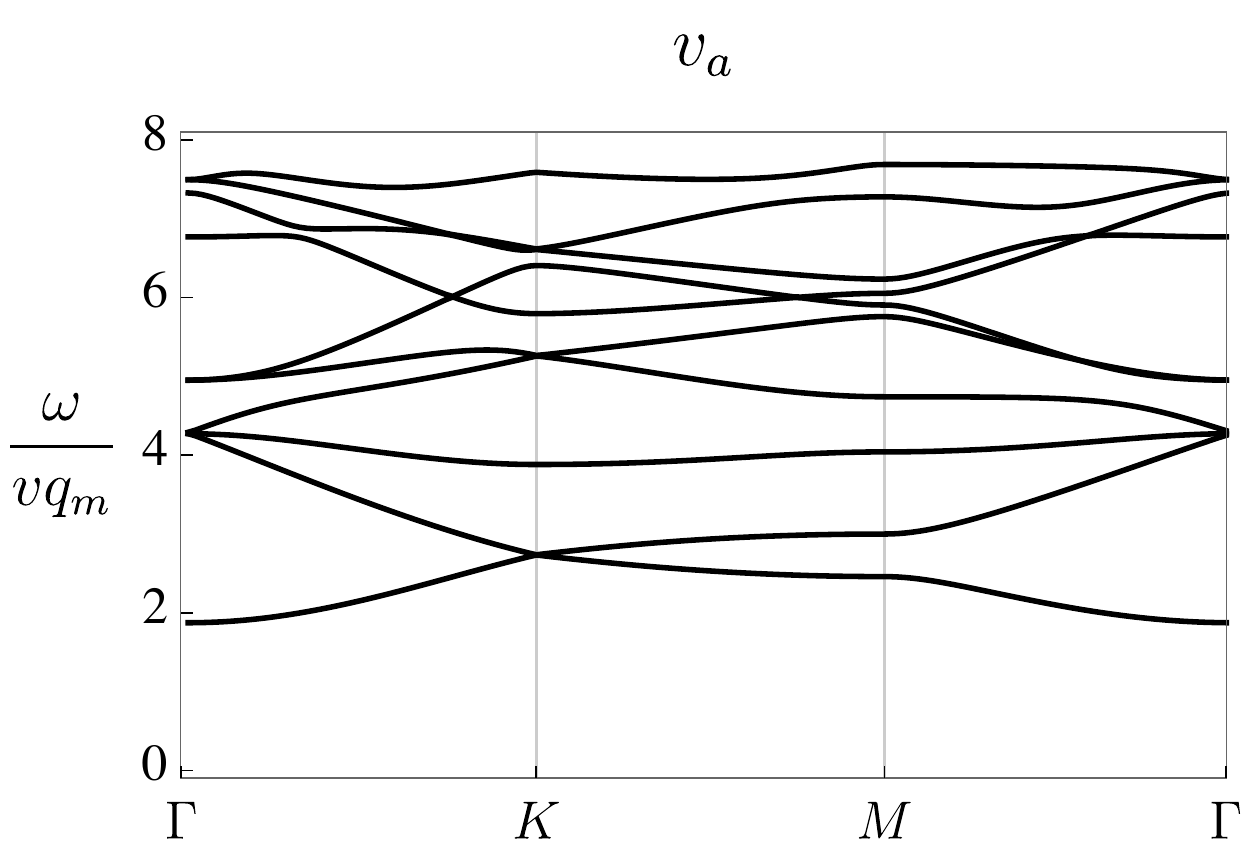}
    \caption{Top row: The ten lowest magnon bands in the collinear phase for the four branches, where we have chosen $\cos\phi_s=\cos\phi_a=-1$. The dimensionless parameters are $\alpha=1,$ $\beta=9.$ Middle: Magnon bands for the two branches $u,~v$ in the twisted-a phase at $\alpha=9$, $\beta=1$. Bottom: Twisted-s phase for the four decoupled branches at $\alpha=2$, $\beta=0.2$.}
    \label{fig:AnisotropicMagnon}
\end{figure}

The twisted-a phase is the only one that involves the interplay between the symmetric and anti-symmetric modes. As can be observed from \eqref{eq:AnisMagnon}, the $u_s,~u_a$ modes are mixed, so are $v_s$ and $v_a$; the four branches thus combine into two, which we will simply label as $u$ and $v$. There is one Goldstone mode in the $v$-branch. In the twisted-s phase, the four equations decouple, and there is again one Goldstone mode corresponding to the out-of-plane rotation $v_s$. 
\begin{equation}
\begin{split}
& \omega^2 \hat{u}_s=-v^2 q_m^2 \left[ (\nabla_{\sf x}+i{\sf k})^2 +\beta \cos\phi_a \right]\hat{u}_s ,\\
& \omega^2 \hat{u}_a=-v^2 q_m^2 \left[ (\nabla_{\sf x}+i{\sf k})^2 -\alpha \hat{\Phi}({\sf x}) \cos\phi_a +\beta\cos\phi_a \right]\hat{u}_a,\\
& \omega^2 \hat{v}_s=-v^2 q_m^2[(\nabla_{\sf x}+i{\sf k})^2 +\frac{1}{4}(\nabla_{\sf x} \phi_a)^2+\frac{\alpha}{2} \hat{\Phi}({\sf x}) (1-\cos\phi_a)-\frac{\beta}{2}(1- \cos\phi_a)]\hat{v}_s,\\
& \omega^2 \hat{v}_a=-v^2 q_m^2[(\nabla_{\sf x}+i{\sf k})^2 +\frac{1}{4}(\nabla_{\sf x} \phi_a)^2-\frac{\alpha}{2} \hat{\Phi}({\sf x}) (1+\cos\phi_a)-\frac{\beta}{2}(1- \cos\phi_a)]\hat{v}_a.
\end{split}  
\end{equation}
The magnon bands in the three phases are shown in figure \ref{fig:AnisotropicMagnon}.

Similar to the isotropic case, the magnon bands flatten at large $\alpha$ due to their confinement in the disconnected domains in a large potential. Below we show an example $\alpha=19,~\beta=9$ in the twisted-a phase.
\begin{figure}[h]
    \centering
    \includegraphics[scale=0.3]{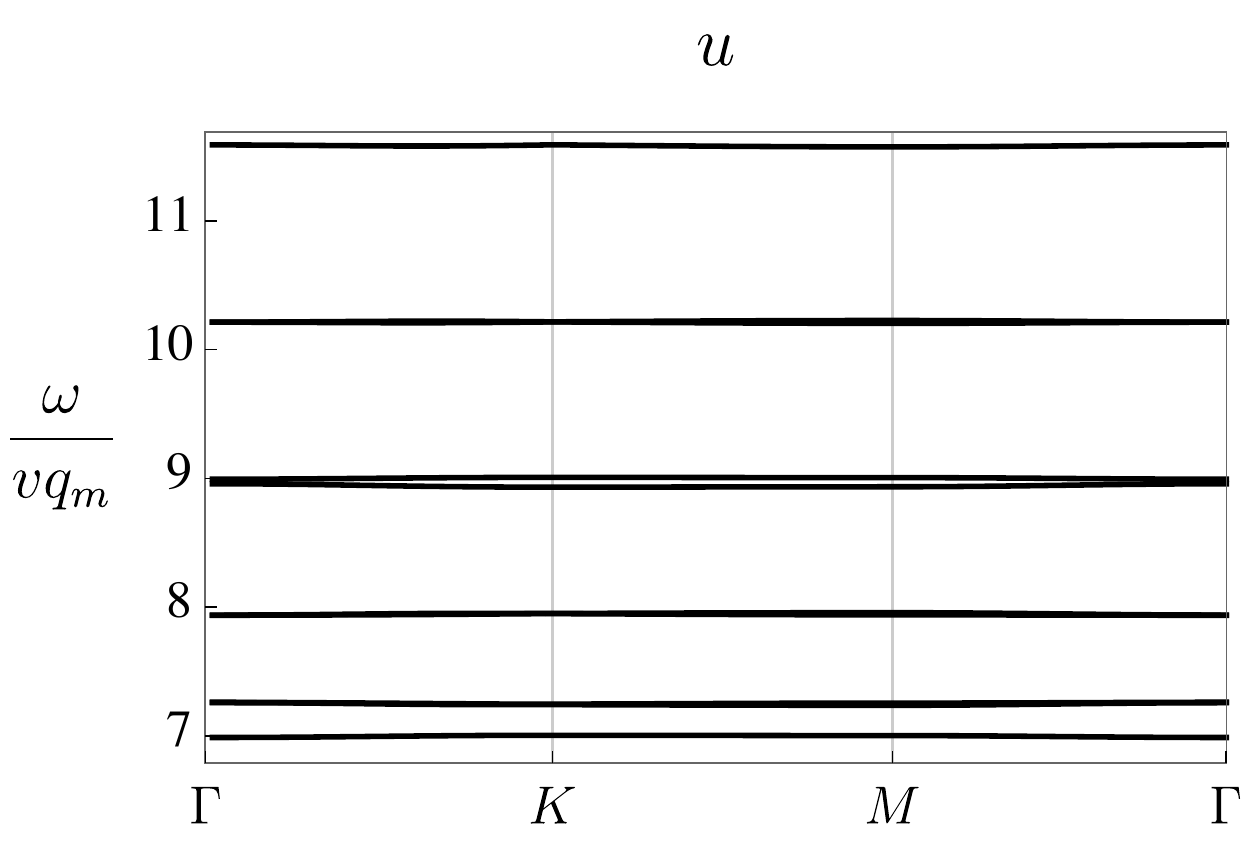}
    \includegraphics[scale=0.3]{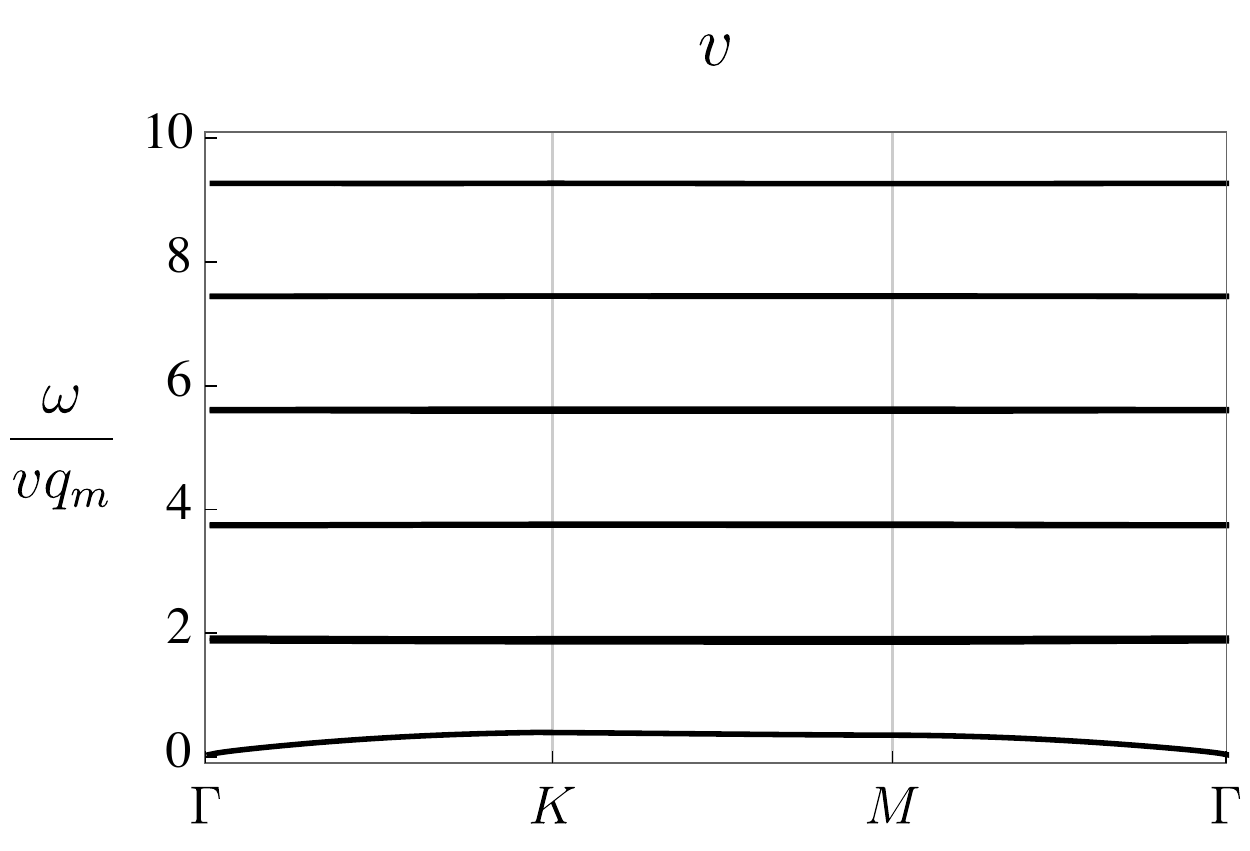}
   \caption{Flattening of magnon bands in the twisted-a phase at $\alpha=19,~\beta=9$.}
    \label{fig:flat}
\end{figure}

\subsubsection*{XY Anisotropy}
Finally, we turn to the case $d<0$, where the Neel vectors tend to lie in the XY plane. We thereby choose the following ansatz:
\begin{equation}
\begin{split}
& \mathbf{N}_l=\sqrt{1-u_l^2-v_l^2}\mathbf{N}_l^{cl}({\sf x})+u_l\mathbf{u}_l({\sf x})+v_l \mathbf{v}_l({\sf x}),\\
\mathbf{N}_l^{cl}=& \sin\phi_l\hat{\mathbf{x}} +\cos\phi_l \hat{\mathbf{y}},\quad \mathbf{u}_l=\cos\phi_l\hat{\mathbf{x}}-\sin\phi_l\hat{\mathbf{y}},\quad \mathbf{v}_l=\hat{\mathbf{z}}.
\end{split}
\end{equation}
The Hamiltonian at the saddle point $\mathbf{N}_l^{cl}$ is ${\sf H}_{cl}=\frac{1}{2}\left[(\nabla_{\sf x} \phi_s)^2+(\nabla_{\sf x} \phi_a)^2\right]-\alpha\hat{\Phi}({\sf x})\cos\phi_a$. Classically, it behaves as if there is no anisotropy, and $\phi_s$ is uniform everywhere. Near the saddle point, following the same procedure as the section above, we obtain in the symmetric/anti-symmetric basis:
\begin{equation}
\begin{split}
{\sf H}_2=&\frac{1}{2}\left[(\nabla_{\sf x} u_s)^2+(\nabla_{\sf x} u_a)^2+(\nabla_{\sf x} v_s)^2+(\nabla_{\sf x} v_a)^2\right]-\frac{1}{8}(v_a^2+v_s^2)(\nabla_{\sf x} \phi_a)^2\\
& + \frac{\alpha}{4}\hat{\Phi}({\sf x}) \left[- v_s^2(1-\cos\phi_a)+v_a^2(1+\cos\phi_a)+2u_a^2\cos\phi_a\right]+\frac{\beta}{2}\left[(v_s^2+v_a^2)\right],
\end{split}
\end{equation}
where the parametrization $\beta=2|d|/\rho q_m^2$ has been used in the last line, which is slightly modified from that in the main text. The corresponding Lagrangian then leads to the following linear wave equations:
\begin{equation}
\begin{split}
& \partial_t^2 u_s=v^2 q_m^2 \nabla_{\sf x}^2 u_s,\quad \partial_t^2 u_a=v^2 q_m^2 \left[ \nabla_{\sf x}^2 u_a -\alpha \hat{\Phi}({\sf x}) \cos\phi_a u_a\right],\\
& \partial_t^2 v_s=v^2 q_m^2[\nabla_{\sf x}^2 v_s +\frac{1}{4}v_s(\nabla_{\sf x} \phi_a)^2+\frac{\alpha}{2} \hat{\Phi}({\sf x}) (1-\cos\phi_a)v_s-\beta v_s],\\
& \partial_t^2 v_a=v^2 q_m^2[\nabla_{\sf x}^2 v_a +\frac{1}{4}v_a(\nabla_{\sf x} \phi_a)^2-\frac{\alpha}{2} \hat{\Phi}({\sf x}) (1+\cos\phi_a)v_s-\beta v_a].
\end{split}   
\end{equation}
All the four branches decouple and almost reduce to the the isotropic form as in the main text, up to the constant shift of $\beta$ in the $v$-branches. Under the Bloch ansatz, the above equations become
\begin{equation}
\begin{split}
& \omega^2 \hat{u}_s=-v^2 q_m^2 \left[ (\nabla_{\sf x}+i{\sf k})^2 \hat{u}_s\right],\\
& \omega^2 \hat{u}_a=-v^2 q_m^2 \left[ (\nabla_{\sf x}+i{\sf k})^2 \hat{u}_a -\alpha \hat{\Phi}({\sf x}) \cos\phi_a \hat{u}_a\right],\\
& \omega^2 \hat{v}_s=-v^2 q_m^2[(\nabla_{\sf x}+i{\sf k})^2 \hat{v}_s +\frac{1}{4}\hat{v}_s(\nabla_{\sf x} \phi_a)^2+\frac{\alpha}{2} \hat{\Phi}({\sf x}) (1-\cos\phi_a)\hat{v}_s-\beta \hat{v}_s],\\
& \omega^2 \hat{v}_a=-v^2 q_m^2[(\nabla_{\sf x}+i{\sf k})^2 \hat{v}_a +\frac{1}{4}\hat{v}_a(\nabla_{\sf x} \phi_a)^2-\frac{\alpha}{2} \hat{\Phi}({\sf x}) (1+\cos\phi_a)\hat{v}_a-\beta \hat{v}_a].
\end{split}  
\end{equation}
There is one Goldstone mode in the $\hat{u}_s$ branch, corresponding to the rotation in the XY plane. As $\alpha$ increases, we will again observe the flattening of $u_a$ and $v_s$ bands. We plot an example in the twisted phase in the figure \ref{fig:XY} below.
\begin{figure}[h]
    \centering
    \includegraphics[scale=0.3]{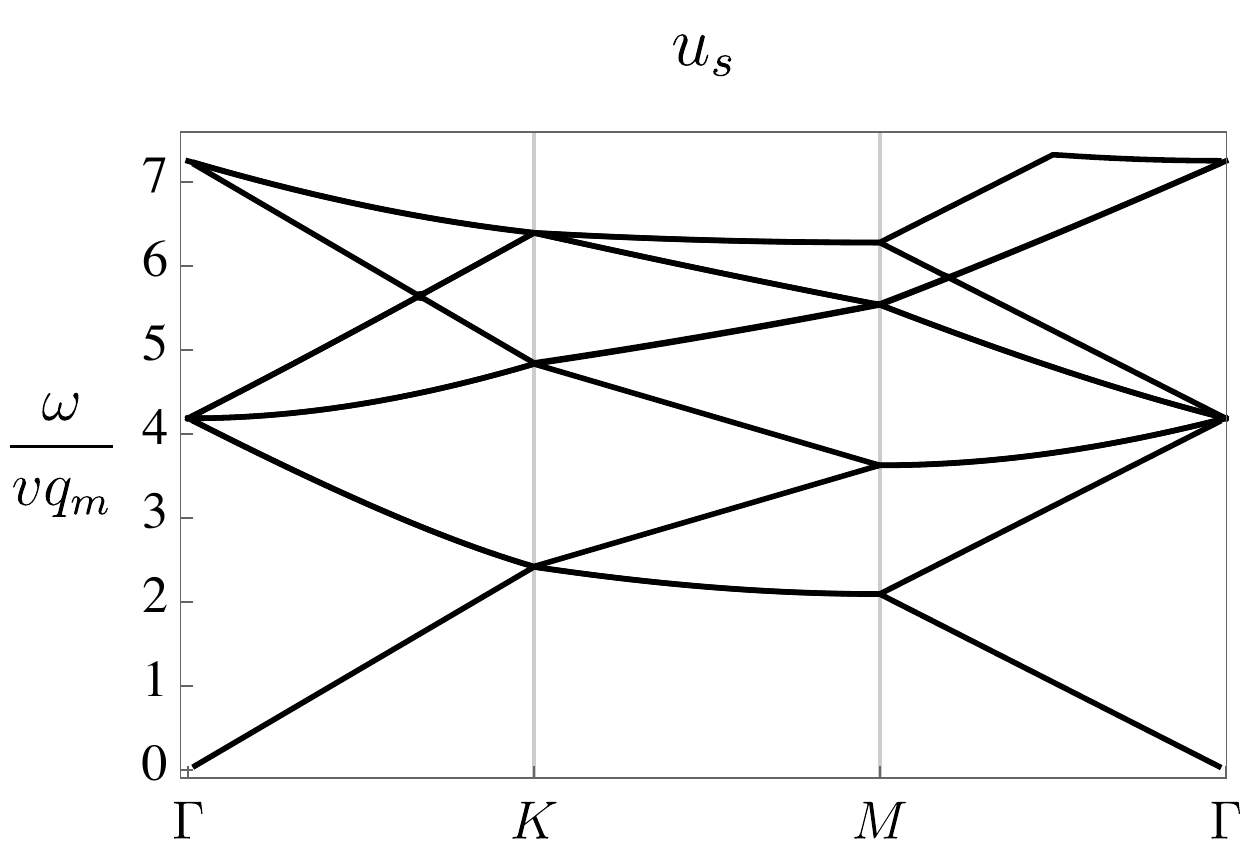}
    \includegraphics[scale=0.3]{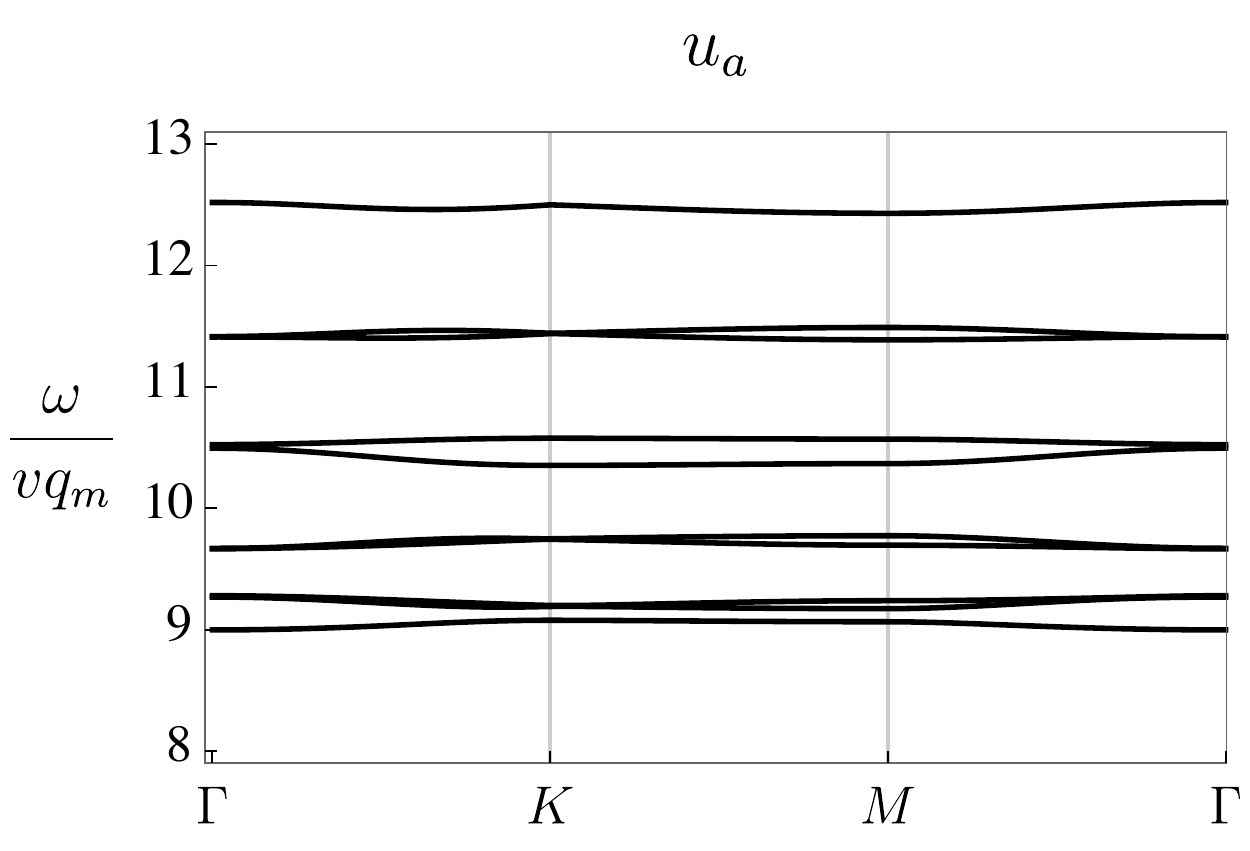}
    \includegraphics[scale=0.3]{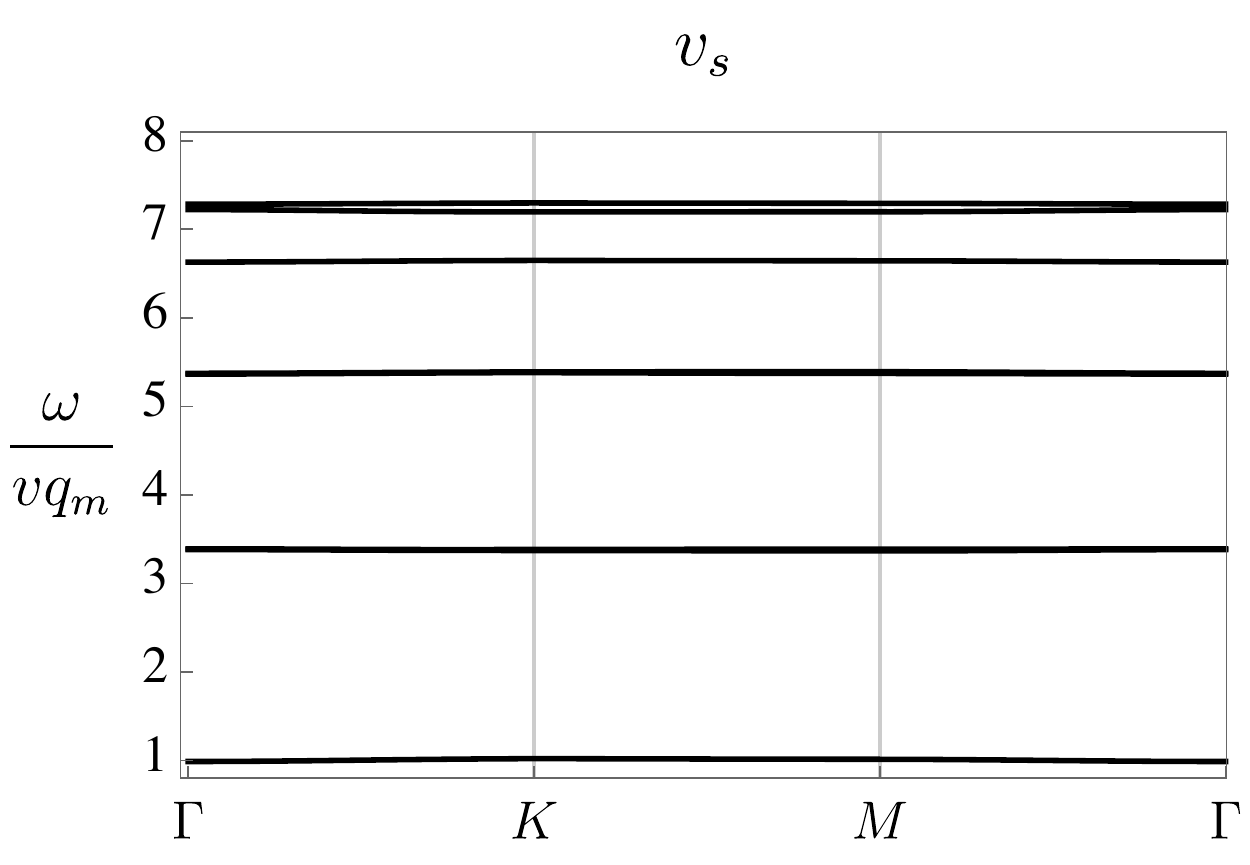}
    \includegraphics[scale=0.3]{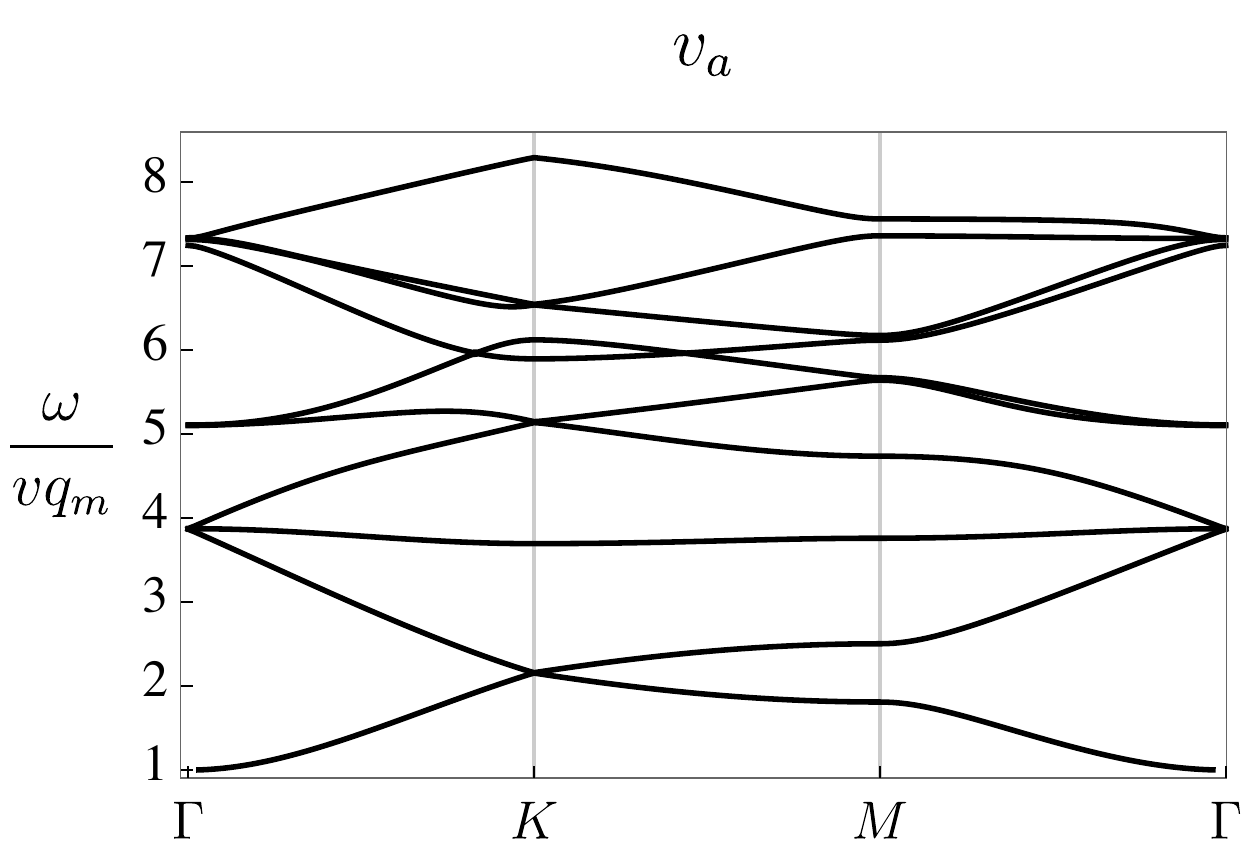}
    \caption{The ten lowest magnon bands for the four branches at $\alpha=9$, $\beta=1$ in the XY-anisotropy case. }
    \label{fig:XY}
\end{figure}

\subsection{General perturbative solution of the Euler Lagrange equations} \label{sec:general_perturbation}

We will encounter the following energy functional and the subsequent partial differential equation in different situations in this work and thus we will first present a general study here. One needs to minimize an energy functional of the following form:
\begin{equation}
  \label{eq:functional_general}
 {\sf H}_{\rm cl} = \frac{1}{2}
   \left|\bm{\nabla}_{\sf x}\phi \right|^2
    - \alpha \left(\xi(\sf{x}) + \xi_0 \right) \cos
  \phi.
\end{equation}
$\xi(\sf{x})$ is a periodic function defining a triangular lattice, and it has zero mean $\int_{\text{unit cell}} \; d^2\sf{x} \; \xi(\sf{x}) = 0$ over one unit cell. One can write a Fourier expansion for $\xi(\sf{x})$ in terms of the reciprocal lattice vectors $\hat{\bm{q}}$ of the above traingular lattice:
\begin{equation}
    \xi({\sf x}) = \sum_{\hat{\bm{q}} \neq 0} \xi_{\hat{\bm{q}}} \; e^{i \hat{\bm{q}} \cdot {\sf x}},
\end{equation}
We will be seeking solutions for $\phi$ that minimize the energy functional above and are periodic with the same period as that given by $\xi({\sf x})$. There is always a trivial solution $\phi = 0$, with an average energy per unit cell equal to $-\xi_0\alpha$. Here we present a perturbative calculation of a nontrivial solution when $\alpha$ and $\xi_0$ are small; these two parameters are taken to be small with their ratio $\frac{\xi_0}{\alpha}$ kept a constant.

In order to find the function $\phi$ that minimizes the above energy functional, we will use the following Euler Lagrange equation:
\begin{equation}\label{eq:euler_lagrange_general}
  \nabla_{\sf x}^2\phi = \alpha \left( \xi({\sf x}) + \xi_0 \right) \sin\phi,
\end{equation}
which should be solved with periodic boundary conditions. Since, we are interested in the specific limit of both $\alpha$ and $\xi_0$ being small, while keeping their ratio $\delta = \frac{\xi_0}{\alpha}$ a constant, we will add a bookkeeping parameter $\varepsilon$ to keep track of orders in our perturbation; we will ultimately set $\varepsilon = 1$. The equation thus takes the form:
\begin{equation}
  \nabla_{\sf x}^2\phi = \varepsilon \; \alpha \left( \xi({\sf x}) + \varepsilon \xi_0 \right) \sin\phi.
\end{equation}
We will find a nontrivial solution as a power series in $\varepsilon$:
\begin{equation}
    \phi = \phi^{(0)} + \varepsilon \, \phi^{(1)} + \varepsilon^2 \,  \phi^{(2)} + \ldots
\end{equation}

To zeroth order in $\varepsilon$, one needs $\phi$ to be a constant, this constant will be determined in higher orders:
\begin{equation}
    \phi^{(0)} = \text{const.} \, .
\end{equation}
To first order in $\varepsilon$, the differential equation takes the form:
\begin{equation}
    O(\varepsilon): \qquad \sum_{\hat{\bm{q}} \neq 0} \left( -\left| \hat{\bm{q}} \right|^2 \right)  \phi^{(1)}_{\hat{\bm{q}}}  e^{i \hat{\bm{q}}  \cdot {\sf x}} = \alpha \sin\phi^{(0)} \sum_{\hat{\bm{q}} \neq 0}   \xi_{\hat{\bm{q}}}  e^{i \hat{\bm{q}}  \cdot {\sf x}}.
\end{equation}
It could be satisfied if $\phi^{(1)}$ takes the form:
\begin{equation}
    \phi^{(1)} = - \alpha \sin \phi^{(0)} \left( \sum_{\hat{\bm{q}} \neq 0} \frac{1}{\left| \hat{\bm{q}} \right|^2} \, \xi_{\hat{\bm{q}}} \; e^{i \hat{\bm{q}}  \cdot {\sf x}} \right) \ +  \ \phi^{(1) \left[\hat{\bm{q}} = 0 \right]},
\end{equation}
where $\phi^{(1) \left[\hat{\bm{q}} = 0 \right]}$ denotes a constant needed in the first order solution, this constant should also be fixed using higher orders of the equation. The second order in $\varepsilon$ of the differential equation now reads:
\begin{equation} \label{eq:diff_eq_o_epsilon_2}
    O(\varepsilon^2): \qquad \sum_{\hat{\bm{q}} \neq 0} \left( -\left| \hat{\bm{q}} \right|^2 \right)  \phi^{(2)}_{\hat{\bm{q}}}  e^{i \hat{\bm{q}}  \cdot {\sf x}} = \alpha \cos\phi^{(0)} \left( \sum_{\hat{\bm{q}}_1 , \hat{\bm{q}}_2 \neq 0}   \xi_{\hat{\bm{q}}_1} \phi^{(1)}_{\hat{\bm{q}}_2}  \; e^{i \left( \hat{\bm{q}}_1 + \hat{\bm{q}}_2 \right) \cdot {\sf x}} \right) + \alpha \xi_0 \sin\phi^{(0)}.
\end{equation}
The left hand side of the above equation does not contain a $\hat{\bm{q}} = 0$ component while the right hand side does:
\begin{equation}
    - \alpha^2 \cos\phi^{(0)} \sin\phi^{(0)} \sum_{\hat{\bm{q}} \neq 0} \frac{1}{\left| \hat{\bm{q}} \right|^2} \, \left| \xi_{\hat{\bm{q}}} \right|^2 + \alpha \xi_0 \sin\phi^{(0)}.
\end{equation}
This needs to vanish so that the second order differential equation holds, and this fixes the value of $\phi^{(0)}$:
\begin{equation}
    \cos\phi^{(0)} = \frac{\xi_0}{\alpha} \frac{1}{\sum \frac{1}{\left| \hat{\bm{q}} \right|^2} \, \left| \xi_{\hat{\bm{q}}} \right|^2} = \delta \; \frac{1}{\sum \frac{1}{\left| \hat{\bm{q}} \right|^2} \, \left| \xi_{\hat{\bm{q}}} \right|^2} .
\end{equation}
This result shows that for values of $\delta$ smaller than $\sum \frac{1}{\left| \hat{\bm{q}} \right|^2} \, \left| \xi_{\hat{\bm{q}}} \right|^2$, a nontrivial solution could exist. 
Furthermore, the second order part of $\phi$ could be found also using \eqref{eq:diff_eq_o_epsilon_2}:
\begin{equation}
\begin{aligned}
    \phi^{(2)} &= \alpha^2 \sin \phi^{(0)} \cos \phi^{(0)} \left( \sum_{\hat{\bm{q}}_1 \neq - \hat{\bm{q}}_2}   \xi_{\hat{\bm{q}}_1} \xi_{\hat{\bm{q}}_2} \frac{e^{i \left( \hat{\bm{q}}_1 + \hat{\bm{q}}_2 \right) \cdot {\sf x}}}{\left|\hat{\bm{q}}_1\right|^2 \; \left|\hat{\bm{q}}_1 + \hat{\bm{q}}_2\right|^2 } \right) \\
    & \qquad - \alpha \cos \phi^{(0)} \left( \sum_{\hat{\bm{q}} \neq 0} \frac{1}{\left| \hat{\bm{q}} \right|^2} \, \xi_{\hat{\bm{q}}} \; e^{i \hat{\bm{q}}  \cdot {\sf x}} \right) \; \phi^{(1) \left[\hat{\bm{q}} = 0 \right]} \ +  \ \phi^{(2) \left[\hat{\bm{q}} = 0 \right]},
\end{aligned}
\end{equation}
with the constant $\phi^{(2) \left[\hat{\bm{q}} = 0 \right]}$ determined again by higher orders of the differential equation.

This procedure can be carried out order by order, we will just state the result for $\phi^{(1) \left[\hat{\bm{q}} = 0  \right]}$, which could be derived from the $\hat{\bm{q}} = 0$ component of the $\varepsilon^3$ order of the differential equation:

\begin{equation}
    \phi^{(1) \left[\hat{\bm{q}} = 0\right]} = \frac{\alpha}{2} \; \frac{ \left( 1 - 3 \cos^2 \phi^{(0)}\right)}{ \sin \phi^{(0)}} 
    \frac{ \sum_{\hat{\bm{q}}_1,\hat{\bm{q}}_3,\hat{\bm{q}}_3} \frac{ 1}{\left|\hat{\bm{q}}_1\right|^2 \; \left| \hat{\bm{q}}_2\right|^2 } \; \xi_{\hat{\bm{q}}_1} \xi_{\hat{\bm{q}}_2} \xi_{\hat{\bm{q}}_3}  \;\delta_{\hat{\bm{q}}_1+\hat{\bm{q}}_2 + \hat{\bm{q}}_3 , 0} }{ \sum_{\hat{\bm{q}}} \frac{1}{\left| \hat{\bm{q}} \right|^2} \left|\xi_{\hat{\bm{q}}} \right|^2 }.
\end{equation}
Also, the average energy density per unit cell can be found to be:
\begin{equation}
    \overline{{\sf H}_{\text{cl}}} = - \frac{\alpha^2}{2} \left( \sum_{\hat{\bm{q}} \neq 0 } \frac{1}{\left| \hat{\bm{q}} \right|^2} \left|\xi_{\hat{\bm{q}}} \right|^2  \right) \left( 1 + \cos^2\phi^{(0)} \right) + O \left( \alpha^3 \right),
\end{equation}
where $O \left( \alpha^3 \right)$ denotes any cubic power of $\alpha$ and $\xi_0$. This energy should be compared with the trivial solution energy density, i.e.~$-\xi_0 \alpha$; the twisted solution, when it exists, has lower energy to this order and thus it is the true ground state for $\delta < \sum \frac{1}{\left| \hat{\bm{q}} \right|^2} \, \left| \xi_{\hat{\bm{q}}} \right|^2$. At $\delta = \sum \frac{1}{\left| \hat{\bm{q}} \right|^2} \, \left| \xi_{\hat{\bm{q}}} \right|^2$, interestingly, the two solutions coincide and thus this transition is continuous.

Finally we would like to emphasize that the above perturbative expansion works when both $\alpha$ and $\xi_0$ are small with their ratio $\delta = \frac{\xi_0}{\alpha}$ kept constant; $\delta$ could be small or order one but the perturbation breaks down for large $\delta$. Below, we will elaborate on the three cases that the above perturbative calculation has been used in this work.

\subsection{Twisted antiferromagnets}

One should consider solving the following Euler-Lagrange equations for a twisted antiferromagnet as discussed in the main text:
\begin{align}
  \nabla_{\sf x}^2 \phi_s & = \beta \cos \phi_a \sin \phi_s, \\
  \nabla_{\sf x}^2\phi_a & = \left(\beta \cos\phi_s +
                       \alpha \hat\Phi({\sf x})\right) \sin\phi_a,
\end{align}
with $\hat\Phi({\sf x}) =  \sum_{a=1}^3 \cos (\bm{\hat{q}}_a \cdot
\sf{x})$ and $\left|\bm{\hat{q}}_a \right| = 1$. One can find a nontrivial twisted solution (which we call twisted-s) in the main text, by setting $\phi_s = 0$ or $\pi$; it will be shown below that the $\phi_s = \pi$ solution has lower energy. Starting from the twsited-s solution, increasing $\beta$ at small $\alpha$ results in a transtion to the collinear phase, while on the other hand for large $\alpha$, with increasing $\beta$, one encounters a transition to the twisted-a phase. We will discuss these two cases separately below.

\subsubsection*{Transtion from the twisted-s phase to the collinear phase, large angles}
At large angles, both $\alpha$ and $\beta$ are small and we will treat the Euler-Lagrange equations perturbatively. With choosing $\beta = 0$ or $\pi$, the equations read:
\begin{equation}
      \nabla_{\sf x}^2\phi_a  = \alpha \left(
                        \hat\Phi({\sf x}) \pm \alpha \; \delta \right) \sin\phi_a,
\end{equation}
where $+$ corresponds to $\phi_s = 0$ and $-$ corresponds to $\phi_s = \pi$, and we will be considering the limit where the ratio $\delta = \frac{\beta}{\alpha^2}$ is kept constant, so that we can use the perturbation series developed above; $\hat\Phi({\sf x})$ plays the role of $\xi({\sf x})$, and $\alpha \ \delta$ plays the role of $\xi_0$.
The $\phi_a$ solution can be found order by order as discussed above:
\begin{equation}
    \phi_a = \cos^{-1} \left(\pm \frac{2}{3} \delta \right) - \alpha \sin\phi^{(0)} \left(  \hat\Phi({\sf x} ) - \left[ \frac12 - \cot^2\phi^{(0)} \right] \right) + O(\alpha^2,\beta).
\end{equation}

The energy density can also be calculated which leads to:
\begin{equation}
    \overline{{\sf H}_{\text{cl}}} = - \frac{3}{4}\alpha^2   \left( 1 + \frac49 \delta^2 \right) \quad \pm \frac16 \alpha^3 \delta \left( 1 + 4 \delta^2 \right) \quad + O \left( \alpha^4 \right).
\end{equation}
This result is kept to one higher order than the previous section; it is this higher order which shows that $\phi_s=\pi$ is preferred energetically and so the $-$ sign should be chosen throughout.

Also, it is worthwhile to note that the limit of $\delta \to 0$, corresponds to $\cos \phi_a^{(0)} = 0$, which simply means that $\bm{N}_1 \cdot \bm{N}_2= 0$ to lowest order in $\alpha$.

\subsubsection*{Transition from twisted-s phase to the twisted-a phase, small angles}
At small angles, where $\alpha$ is large but $\beta$ is kept still small, one can take the configuration of $\phi_a$ to be completely determined by satisfying the $- \alpha \hat\Phi( {\sf x} ) \cos\phi_a$ term in the Hamiltonian: $\cos\phi_a$ can be taken equal to $\mathrm{sign} \left[ \hat\Phi(\sf x) \right]$. This forms domains of constant $\phi_a$, with narrow domain walls between them. On the other hand, $\phi_s$ should be found using the Euler-Lagrange equations, which reduce to:
\begin{equation}
     \nabla_{\sf x}^2 \phi_s = \beta \cos \phi_a \sin \phi_s.
\end{equation}
For small $\beta$, we can use the perturbation theory developed above, with $\beta$, and $\cos \phi_a = \mathrm{sign} \left[ \hat\Phi(\sf x) \right]$ (remember that $\phi_a$ is not dynamical in the above equation) playing the roles of $\alpha$, and $\xi({\sf x}) + \xi_0$ respectively in \eqref{eq:euler_lagrange_general}. One can see that a twisted solution with $\cos\phi_s^{(0)} = \frac{\xi_0}{\sum \frac{1}{ \left| \hat{\bm{q}} \right|^2} \left| \xi_{\hat{\bm{q}}} \right|^2  } \frac{1}{\beta}$ exists, if $\beta$ is large enough. It is found numerically that 
$$\xi_0 = \frac{1}{A_{\text{u.c.}}}\int_{\text{unit cell}} \mathrm{sign} \left[ \hat\Phi({\sf x}) \right] = -0.21,$$
which means that domains with antiferromagnetic interlayer exchange have larger area than those with ferromagnetic interlayer coupling. Furthermore, one can also find numerically that
$$\sum \frac{1}{ \left| \hat{\bm{q}} \right|^2} \left| \xi_{\hat{\bm{q}}} \right|^2  = 0.71.$$

These two values show that a twisted solution for $\phi_s$ could appear if $\beta > 0.29$; this should correpsond to the $\beta$ value for which the transition between twisted-s and twisted-a phases occurs at large $\alpha$, and it is indeed very close, with a few percent error actually, to the value found numerically at large $\alpha$ in the phase diagram presented in the main text.

\bigskip

\subsection{Twisted ferromagnetic CrI$_3$ bilayer}

For the properties of the interlayer exchange parameter in a bilayer CrI$_3$ system, we will be following the numerical results presented in Ref.~\cite{sivadas2018stacking}, where first-principles calculations are carried out: it is shown that the interlayer exchange can vary considerably if the bilayer stacking is altered, and in fact it can change its sign; the pristine CrI$_3$ bilayer exhibits antiferromagnetic interlayer exchange, but the above statement implies that this can be modified if the stacking is varied. Remarkably in a twisted bilayer, the displacement between the layers is modulated periodically with a unit cell given by the moir\'e length and so all the different kinds of displaced bilayer stacking are realized. With this in mind, one can use the energy functional discussed in the main text
\begin{equation}
  \mathcal{H}_{\rm cl} = \sum_l \left[ \frac{\rho}{2} \left(
      \nabla\bm{M}_l\right)^2 - d
    \left(N_l^z\right)^2\right] - J' \Phi(\bm{x}) \bm{M}_1\cdot \bm{M}_2,
\end{equation}
where, as discussed in the main text, the function $\Phi(\bm{u}_1 - \bm{u}_2)$ takes a form that should be extracted from the calculations of the stacking dependence of interlayer exchange as obtained for example in Ref. \cite{sivadas2018stacking}.

\begin{figure}[b]
    \centering
    \includegraphics[scale=0.4]{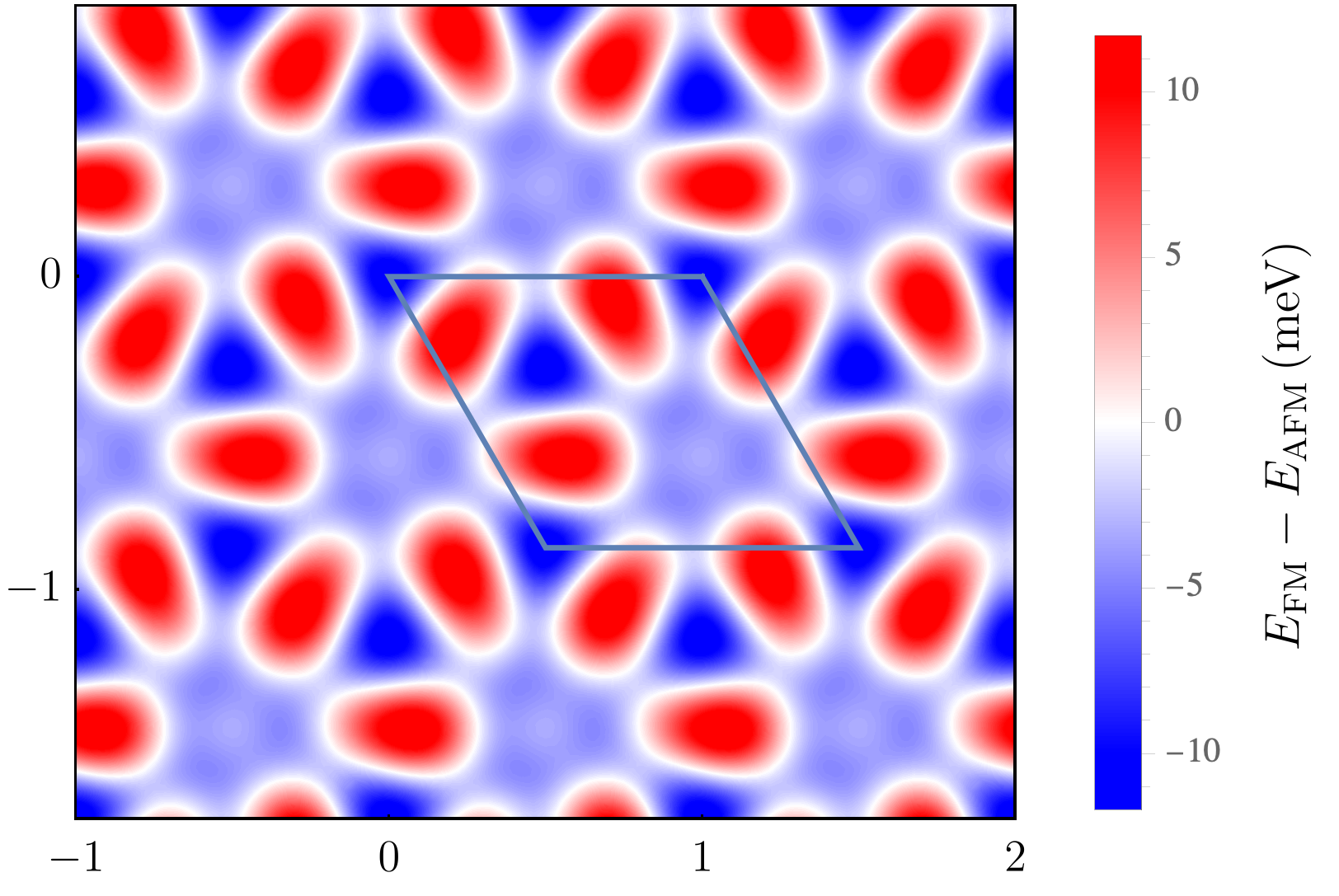}
    \caption{A reproduced plot of interlayer exchange energy per unit cell of bilayer CrI$_3$ as a function of the displacement between the two layers. The data is extracted from figures in Ref.~\cite{sivadas2018stacking}. 
    The two axes show displacement in the two directions in the units of a real space unit cell length; the interlayer exchange is indeed a periodic function.
    The blue and red regions show ferromagnetic and antiferromagnetc interlayer exchange.}
    \label{fig:exchange_cri3}
\end{figure}

We have used the plots presented in Ref. \cite{sivadas2018stacking}, to find the Fourier components of the interlayer exchange which is indeed a periodic function of the interlayer displacement. It turns out that unlike the antiferromagnetic case initially studied in the main text, i.e.~$\Phi(\bm{x}) = \sum_{a=1}^3 \cos (\bm{q}_a \cdot \bm{x})$, the present $\Phi(\bm{u})$ function needs several harmonics along with a constant term to be reproduced (see Fig.~\ref{fig:exchange_cri3}). We will ultimately work with a rescaled Hamiltonian that has a form that is identical to that in the antiferromagnetic case:
\begin{equation}
  \label{eq:hamiltonain_FM}
 {\sf H}_{\rm cl} = \frac{1}{2}
  \left( |\bm{\nabla}_{\sf x}\phi_s|^2 +
    |\bm{\nabla}_{\sf x}\phi_a|^2\right) - (\alpha \hat\Phi(\sf{x})+\beta\cos\phi_s) \cos
  \phi_a.
\end{equation}
$\alpha$ and $\beta$ are defined as before and $\hat{\Phi}({\sf x} ) = \hat{\Phi}_0 + \sum_{\hat{\bm{q}} \neq 0} \hat{\Phi}_{\hat{\bm{q}}} \, e^{i \hat{\bm{q}} \cdot \sf{x} }$, where $\hat{\bm{q}}$'s are the rescaled moir\'e reciprocal lattice vectors. We have kept the lowest five harmonics along with the constant term here. Furthermore, $\hat{\Phi}$ is normalized in a way that $\sum_{\hat{\bm{q}} \neq 0} \frac{1}{|\hat{\bm{q}}|^2}\left| \hat{\Phi}_{\hat{\bm{q}}} \right|^2 = 1.$ Variation of this energy functional leads to the same set of equations
\begin{align}
  \nabla_{\sf x}^2 \phi_s & = \beta \cos \phi_a \sin \phi_s, \\
  \nabla_{\sf x}^2\phi_a & = \left(\beta \cos\phi_s +
                       \alpha \hat\Phi({\sf x})\right) \sin\phi_a,
\end{align}
which should be solved to minimize the energy here also. We will only discuss the case of a positive infinitesimal $\beta$ here, the case of general positive $\beta$ should be similar to the antiferromagnetic case. The effect of a positive infinitesimal $\beta$ is to fix a value for $\phi_s$, and in this case it turns out that $\phi_s = 0$ is energetically favored;  this will be justified below.

The only functionality that needs to be determined now is that of $\phi_a$, and the only parameter is $\alpha$; this time $\hat{\Phi}(\sf{x})$ has a nonzero constant term $\hat{\Phi}_0$ as well, and thus at small $\alpha$ the $\phi_a$ configuration is totally controlled by this constant term; we have derived $\hat{\Phi}_0 = 0.026 > 0$ which means that this constant imposes ferromagetic interlayer exchange, and thus the solution at small $\alpha$ turns out to be $\phi_a = 0$, with an energy density  $\overline{{\sf H}_{\rm cl}} = -\alpha \hat{\Phi}_0$. One expects this trivial state to give way to a twisted solution with lower energy at some value of $\alpha$;  since $\hat{\Phi}_0$ is small itself, the transition to a twisted phase happens at a small $\alpha$ and thus one can set up a perturbative calculation for the twisted solution of $\phi_a$ at small $\alpha$; this perburbative calculation, which is discussed in Sec.~\ref{sec:general_perturbation} of the SM, yields
\begin{equation}\label{eq:phi_a_ferro}
   \phi_a = \phi_a^{(0)}  + \alpha \left( -\sin\phi_a^{(0)}  \sum_{\hat{\bm{q}} \neq 0} \frac{\hat{\Phi}_{\hat{\bm{q}}}}{\left| \hat{\bm{q}} \right|^2}  \, e^{i \hat{\bm{q}} \cdot {\sf x} } +  \phi_a^{(1)[\hat{\bm{q}} = 0]} \right)
   \quad + O(\alpha^2,\alpha \, \hat{\Phi}_0), 
\end{equation}
with $\cos \phi_a^{(0)} = \frac{1}{\alpha}\hat{\Phi}_0$. This means that the twisted solution exists for $\alpha$ above $\alpha_0 = \hat{\Phi}_0 = 0.026$ to leading order. The energy density for this state turns out to be $\overline{{\sf H}_{\rm cl}} = -\frac12 \left( \alpha^2 + \hat{\Phi}_0^2 \right)$ to leading order; this shows that indeed a continuous transition to the twisted phase happens at $\alpha = \alpha_0$.

For very large values of $\alpha$ similar to what happens in the twisted antiferromagnets discussed in the main text, the twisted solution implies that $\phi_a$ is either $0$ or $\pi$ almost everywhere, so that $\cos\phi_a = \mathrm{sign}[\hat{\Phi}(\sf{x})]$ except for narrow domain wall regions where $\hat{\Phi}(\sf{x})=0$. 

Here we can see why $\phi_s = 0$ is chosen for an infinitesimal positive $\beta$ in two different limits: at small $\alpha$, the constant term $\hat{\Phi}_0$ is ferromagnetic and thus the energy will decrease by setting $\phi_s=0$; for large $\alpha$ on the other hand, since one is in the extreme twisted phase, one should note that the area with ferromagnetic interlayer coupling is larger than that with antiferromagnetic coupling, or in other words 
$$\frac{1}{A_{\text{u.c.}}}\int_{\text{unit cell}} \mathrm{sign} \left[ \hat\Phi({\sf x}) \right] > 0,$$
and thus $\phi_s = 0$ is again energetically favored. It is worthwhile to mention that this is a coincidence in CrI$_3$, that both small and large $\alpha$ limits prefer interlayer ferromagnetism; this could well not be the case in other materials in which cases it is reasonable to expect a transition at intermediate $\alpha$ from $\phi_s = 0$ to $\phi_s = \pi$.

\FloatBarrier

\bibliography{moire}

\end{document}